\documentstyle[12pt,axodraw,epsf,epsfig]{article}
\newcommand{\citeup}[1]{$\!^{\mbox{\scriptsize \cite{#1}}}$}

\begin{document}
\begin{flushright}
MC-TH-98-17\\
October 1998\\
Last revised: January 1999\\
\end{flushright}
\begin{center}
\vspace*{2cm}

{\Large \bf New explanation of the
strange baryon rapidity distributions in ultra-relativistic 
nucleus-nucleus collisions.} \\

\vspace*{1cm}

J. A. ~Casado
\footnote{
E-mail: {\tt casado@a13.ph.man.ac.uk}
}

\vspace*{0.5cm}
Department of Physics and Astronomy,\\
University of Manchester,\\
Manchester, M13 9PL, England.

\end{center}
\vspace*{5cm}
    
\begin{abstract}

A model of multiparticle production in hadronic collisions at 
ultra-relativistic energies, based on the assumption of independent 
string fragmentation, reproduces the rapidity spectra of $\Lambda$ 
and $\bar{\Lambda}$ in sulphur-sulphur collisions reported by the 
NA35 Collaboration. This is achieved after a reconsideration of the 
intermediate multi-string states and the structure of the diquarks.  
Nuclear stopping power is also studied through the computation of 
the $p-\bar{p}$ rapidity spectra.

\end{abstract}

\newpage

\section{Introduction}
Strange particle production in heavy ion collisions is a 
subject of great interest since it may provide evidences of the 
formation of a new state of matter: the quark-gluon plasma 
(QGP). The tantalizing pursuit  of this discovery is the justification 
for the efforts being made to carry out the heavy-ions collision 
experiments that have taken place at the AGS and CERN-SPS, and to plan 
new ones for future facilities. A phase transition 
between normal nuclear matter to a deconfined phase has been 
predicted long ago\citeup{CollinsQGP,PolyakovQGP} to occur at very high 
temperatures and densities. For the time being, the properties of 
the QGP have not being clearly stablised, neither the order of the 
phase transition between the two states. This makes the problem of 
recognizing the formation of QGP, if it happens, particularly difficult, 
more so since we have to rely on indirect clues to its detection. 
Several signatures have being proposed for this purpose  
that may deliver circumstantial evidences of the formation of quark 
matter (for an up-to-date review on the subject with a 
complete list of references, see \cite{MiklosQGP}).
The strange mesons, baryons and antibaryons production has been 
predicted to be greatly enhanced in the presence of QGP as compared 
to the case when there are only standard hadron matter\citeup{Rafelski,Rafelski2}. 
Also the study of the stopping power may help to uncover the presence of the 
deconfined phase: a rapid change in the shape of the rapidity baryon 
distributions with increasing incident beam energy is thought to be a 
clear signal for new degrees of freedom\citeup{MiklosStopping}. 

An additional problem is caused by the lack of a full description of 
the confined phase, which constitutes the background to the 
events in which QGP may have been formed. {\em Enhanced strange particle 
production} is a meaningless expression unless we can find out what 
strange particle production must be expected from an ordinary ion-ion 
collision event. We can not extract this 
information from QCD since we do not know how to apply this theory when the 
effective coupling constant is not small. This is also the ultimate cause of 
the lacunae in our understanding of the QGP properties and of the 
order of the phase 
transition  mentioned in the above paragraph. We have to rely on 
phenomenological models to fill this gap. The Dual Parton Model 
(DPM)\citeup{DPMPhysRep} 
and the Quark Gluon String Model (QGSM)\citeup{QGSM1,QGSM2} 
have been very successful in 
describing many features of hadron-hadron, hadron-nucleus and 
nucleus-nucleus collisions. Nevertheless, they were not able to 
reproduce\citeup{CapellaMerinoStrange} 
the rapidity distributions 
of $\Lambda$ and $\bar{\Lambda}$ hyperons, as well as proton minus 
antiprotons ($p-\bar{p}$), in central 
sulphur-sulphur collisions measured by the NA35 
Collaboration\citeup{NA35Lambda,NA35Proton}.
The idea that the onset of the deconfinement phase transition may be at 
the origin of this disagreement is very appealing. However, we must 
carefully explore any other possible explanation of these data before 
admitting that eventuality. The NA35 data show a large production 
of $\Lambda$'s in the fragmentation as well as  in the central region. 
It is very difficult to understand why the formation of QGP in a collision 
between identical nuclei should contribute so largely to the 
borders of the accessible phase space. This can not obviously be due to 
QGP formation as it should mainly affect the hadron spectra in 
the central rapidity region. Besides, QGP formation in this 
experiments, involving relatively light nuclei, would come 
as a striking surprise. The energy 
densities currently available at the  AGS and CERN-SPS of 0.5-10 
$\mbox{GeV}/\mbox{fm}^{2}$ correspond to temperatures between 100 
and 200 MeV\citeup{Bjorken}, for which the effective coupling 
constant of QCD is not less than 1, and we should not be 
expecting a deconfinement phase transition to occur. This also 
should encourage us to look for conventional explanations to the 
phenomena observed in heavy nucleus collisions that have not 
been understood so far. 

Strangeness enhancement is also present in $pp$ interactions. 
The ratio $K/\pi$ increases both with the energy of the collision and 
with the multiplicity\citeup{E755}. It has also been observed an
increase of the ratio $\bar{\Lambda}/\bar{p}$ in $pp$ 
collisions as the centre-of-mass energy increases from 20 GeV to 1800 
GeV\citeup{ExpHypHE}. Again, the NA35 Collaboration also reported that 
strangeness is enhanced between $pp$ and $pS$ collisions\citeup{NA35Lambda}.
At the same time, the average number of collisions per 
participant nucleon increases between $pp$ and $pA$ collisions as 
well as between $pA$ and central $AB$ collisions. This increase is also 
correlated with an increase of the multiplicity of secondaries. According 
to the DPM and the QGSM, the extra particles are produced by the 
fragmentation of strings that have sea quarks at their ends. As this 
mechanism on its own failed to reproduce the observed enhancement 
of $\Lambda$ and $\bar{\Lambda}$ in central $SS$ collisions, 
additional sources of this increase have been explored. 

Diquark-antidiquark pairs from the nucleon sea have been 
introduced\citeup{Ranft1,Ranft2,DiqCapella}. 
This model predicts, contrary to the experimental evidence, the same 
increase in the absolute value of strange baryons and antibaryons.  It may 
also be worth mentioning that this mechanism is very similar in its 
formulation and its results to the, so called, string 
fusion model of ref. \cite{StringFusion}. In reference 
\cite{CapellaMerinoStrange} the  problem is treated 
considering final state interactions of co-moving pions and nucleons 
of the type $\pi + N \rightarrow  K + \Lambda$. This approach is not 
convincing since it contains the unrealistic assumption that all baryons are 
created instantaneously at the beginning of the collision so the 
secondary collisions can take place.

What the data on strange baryon production of the NA35 Collaboration 
may actually be suggesting is that we may be missing an important 
piece in our understanding of the nucleon-nucleon rescattering. I 
devote this paper to reconsider how the 
valence and sea quarks take part in the formation of 
the ends of strings. I will show that, by allowing the leading 
diquarks to be formed not only of valence quarks, a good 
understanding of the data in the fragmentation region  is obtained. 
I also considered the net proton distribution, usually regarded as a 
measurement of the stopping power of nuclear matter. Data form the 
NA35 Collaboration shows a larger stopping than predicted. The 
model proposed here also gives better results for this particular 
problem.

There have been a couple of other mechanisms proposed that could 
contribute to the full explanation of the data. One of 
them\citeup{DB} assumes 
that diquarks can break up into two quarks, one of which can be slowed 
down along the string carrying the string junction (baryon number) 
and produces much slower baryons. The other\citeup{kharzeev} 
consists in considering 
the possibility that the baryon number is not always carried by a 
diquark, or a quark as in the previous case, but by a non-perturbative 
configuration of gluon fields. Both have similar phenomenological 
implications as both mainly affect the centre of the rapidity space 
and can not account for what happens in the fragmentation region what 
is my major concern in this paper, in particular for the case of 
$\Lambda$ production.

There are a number of other models that are also constructed 
around the idea of independent fragmentation of strings. 
The best known is the Lund-Fritiof model\citeup{Fritiof}. Although the 
general assumptions are at first sight very similar to the 
ones in DPM and QGSM, the theoretical basis from which these models 
were developed are quite different and I will not refer to them in 
the rest of the paper. 

In section \ref{sec:TheModel} I discuss the 
theoretical basis of the paper and make a full 
description of the  model. The next two sections deal with fits to 
$pp$ data to obtain the value of some parameters of the model, and the 
results for nucleus-nucleus collision. The paper ends 
with the conclusions.

\section{The Model} \label{sec:TheModel}
The independent-fragmentation string models, DPM and QGSM, were 
inspired by the developments of the Dual Topological 
Unitarization (DTU)\citeup{DTU1,DTU2,DTU3,DTU4,DTU5}.  
The amplitudes of high-energy hadronic 
collisions  were formally written as an expansion in powers of 
$1/N_{f}$, where $N_{f}$ is the number of flavours. Each term in 
the expansion can be associated with a diagram with a definite 
topology. The scheme was based on the S-matrix analysis and was 
closely related to the Regge Field Theory (RFT).  The DTU program was 
reinforced when Veneziano\citeup{Veneziano} established a 
conceptual link with QCD 
as he  included  in the series powers of $1/N_{c}$\citeup{NC1,NC2,NC3}, 
being $N_{c}$ 
the number of colours. In  this way, the whole series of the 
topological expansion formally includes all the terms of the 
perturbative QCD series ensuring unitarity.

\begin{figure}[t]
\begin{center}\begin{picture}(300,140)(0,-20) 
\Line(0,0)(120,0)
\Line(0,100)(120,100)
\ZigZag(60,0)(60,100){10}{5}
\DashLine(60,-10)(60,110){10}
\Text(30,50)[]{$R$}
\Text(130,50)[]{$=$}
\Text(10,90)[]{$\pi$}
\Text(10,10)[]{$\pi$}
\ArrowLine(160,100)(260,100)
\ArrowLine(260,0)(160,0)
\CArc(170,80)(10,0,90)
\Line(160,90)(170,90)
\ArrowLine(180,20)(180,80)
\CArc(170,20)(10,-90,0)
\Line(160,10)(170,10)
\ArrowArc(260,80)(10,90,270)
\ArrowArc(260,50)(10,90,270)
\ArrowArc(260,20)(10,90,270)
\Line(145,0)(145,100)
\Text(155,95)[]{$\pi$}
\Text(155,5)[]{$\pi$}
\Text(270,95)[]{$h_1$}
\Text(270,65)[]{$h_2$}
\Text(270,35)[]{$h_3$}
\Text(270,5)[]{$h_4$}
\Line(280,0)(280,100)
\Text(290,100)[]{{\Large 2}}
\end{picture} \end{center}
\caption{Representation of a cut reggeon and its 
relation with the planar diagram.} \label{fig:OneReggeon}
\end{figure}
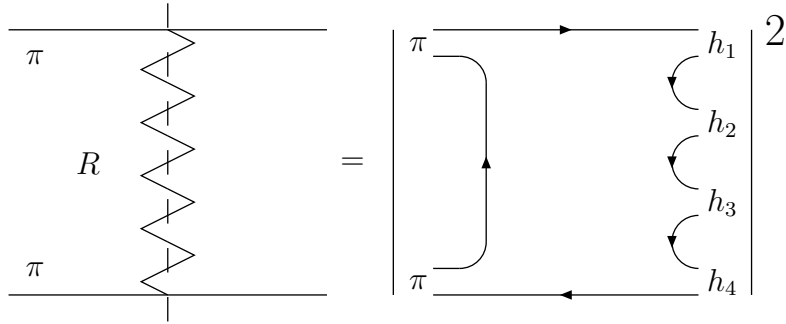

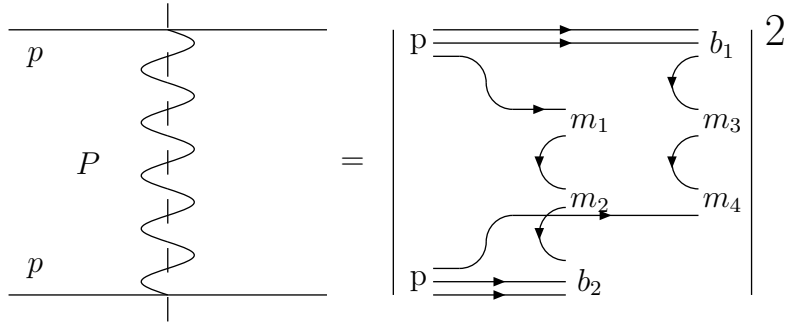
\begin{figure}[t] 
\begin{center}
\begin{picture}(300,140)(0,-20) 
\Line(0,0)(120,0)
\Line(0,100)(120,100)
\Photon(60,0)(60,100){10}{5}
\DashLine(60,-10)(60,110){10}
\Text(30,50)[]{$P$}
\Text(130,50)[]{$=$}
\Text(10,90)[]{$p$}
\Text(10,10)[]{$p$}
\ArrowLine(160,100)(260,100)
\ArrowLine(160,95)(260,95)
\ArrowLine(160,5)(210,5)
\ArrowLine(160,0)(210,0)
\CArc(170,80)(10,0,90)
\Line(160,90)(170,90)
\CArc(190,80)(10,180,270)
\ArrowLine(190,70)(210,70)
\CArc(170,20)(10,-90,0)
\Line(160,10)(170,10)
\CArc(190,20)(10,90,180)
\ArrowLine(190,30)(260,30)
\ArrowArc(260,80)(10,90,270)
\ArrowArc(260,50)(10,90,270)
\ArrowArc(210,50)(10,90,270)
\ArrowArc(210,23)(10,90,270)
\Line(145,0)(145,100)
\Text(155,95)[]{p}
\Text(155,5)[]{p}
\Text(270,95)[]{$b_1$}
\Text(270,65)[]{$m_3$}
\Text(270,35)[]{$m_4$}
\Text(220,5)[]{$b_2$}
\Text(220,65)[]{$m_1$}
\Text(220,35)[]{$m_2$}
\Line(280,0)(280,100)
\Text(290,100)[]{{\Large 2}}
\end{picture}
\end{center}
\caption{Diagrammatic representation of a cut pomeron and its 
relation with the cylinder diagram.} \label{fig:OnePomeron}
\end{figure}

The lowest order term (fig. \ref{fig:OneReggeon} ) is the 
planar diagram connecting two colliding-hadron lines. 
It represent the exchange of an object that has the same 
quantum number of the Reggeon and can be identified with it.
Unitarity requires the inclusion of all diagrams obtained 
by repeatedly multiplying the planar term by itself. The 
next order topology we have to consider is that of the cylinder 
that corresponds to the Pomeron (see fig. \ref{fig:OnePomeron}) 
as it only vacuum quantum numbers can be exchange in the elastic
collision.
Higher order terms can be obtained by combining the simplest 
topologies described above, and all of them have a one to one 
correspondence with Reggeon Field Theory (RFT) diagrams. Bearing 
in mind that we are interested in the high-energy regime,  we make 
use of this correspondence to neglect terms containing the exchange 
of one or more reggeons: although they may be dominant in the 
topological expansion, their contribution decreases with the 
centre-of-mass energy squared $s$ as $s^{\alpha_{R}(0)-1} \sim 
s^{-0.5}$ while the pomeron exchange subdiagrams behave 
like $s^{\alpha_{P}(0)-1} = s^{\Delta} $, where $\Delta \geq 0$.

This formal developments laid the foundation of phenomenological 
models that had to fill the conceptual gaps left by the theory. 
Both DPM and QGSM postulated that multiparticle production in 
hadronic collisions at ISR energies, take place in two definite 
steps. First, the hadrons split in two fragments that share the 
total momentum of the hadron according to the corresponding 
structure functions. One of the fragments belongs to the $3$ 
representation of the $SU(3)$ colour group, a quark, and the 
other to the $\bar{3}$ representation, an antiquark in the case 
of a meson or a diquark in the case of a baryon. Secondly, two 
colour strings are formed each one having a fragment of 
the $3$ representation in one end from one of the hadrons, and 
another of the $\bar{3}$ representation, from the other hadron 
at the other end. These two strings decay into two chains of 
hadrons that form the final multiparticle state. The contribution 
of a string to the final state is given by the fragmentation 
functions that are, along with the structure functions,  the 
other  basic ingredient of the models.

When we move to higher energies, we have to introduce terms with more 
than two strings that correspond to diagrams with several cut 
pomerons. The probability weights of the different diagrams are 
given by the perturbative RFT that can be effectively described by a 
generalized eikonal model\citeup{DPMPhysRep}. Each cut pomeron 
add two chains of hadrons to the final state as the result of the 
decay of two more strings. The generalization of the models to 
hadron-nucleus and nucleus-nucleus collisions carries some 
similarities with this extrapolation to higher energies. In this 
case, multiple inelastic collision appears as a consequence of the 
rescattering of each nucleon by the nucleons on the other interacting 
nucleus. Again, for each additional inelastic collision we have 
to consider the fragmentation of two extra chains. The probability 
weights are given in this case by an external model like that of 
Glauber-Gribov, that reproduces the geometry of the collision.

Both DPM and QGSM assume that the diquarks at the end of the strings in 
hadronic collisions, are always made of valence quarks. So, in the 
case of a colliding proton, regardless of the number of cut pomerons 
in the collision, the diquark can only be of type $(uu)$ or $(ud)$. The 
central point to this paper is the assumption that the sea quarks can 
be part of the diquark with the same probability as the valence quarks. 
It is still true that the diquark in a proton can only be of the 
types 
mentioned above when it suffers only one inelastic collision. But, 
when the number of cut pomerons is greater than one, we also have 
to consider diquarks of the type $(dd)$, $(us)$, $(ds)$ and $(ss)$. 
The flavour content of the diquark depends on the flavour content of 
the hadron sea. 

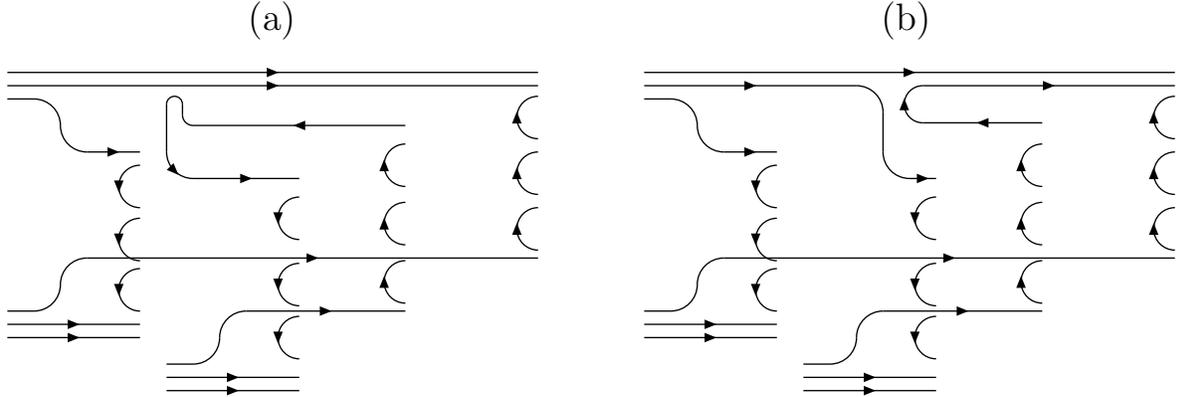
\begin{figure}[t] 
\begin{center}
\begin{picture}(450,160)(-20,0) 
\ArrowLine(0,100)(200,100)
\ArrowLine(0,95)(200,95)
\ArrowLine(0,5)(50,5)
\ArrowLine(0,0)(50,0)
\CArc(10,80)(10,0,90)
\Line(0,90)(10,90)
\CArc(30,80)(10,180,270)
\ArrowLine(30,70)(50,70)
\CArc(10,20)(10,-90,0)
\Line(0,10)(10,10)
\CArc(30,20)(10,90,180)
\ArrowLine(30,30)(200,30)
\ArrowArcn(200,83)(8,270,90)
\ArrowArcn(200,62)(8,270,90)
\ArrowArcn(200,41)(8,270,90)
\ArrowArc(50,57)(8,90,270)
\ArrowArc(50,37)(8,90,270)
\ArrowArc(50,18)(8,90,270)
\ArrowLine(60,-15)(110,-15)
\ArrowLine(60,-20)(110,-20)
\CArc(70,0)(10,-90,0)
\Line(60,-10)(70,-10)
\CArc(90,0)(10,90,180)
\ArrowLine(90,10)(150,10)
\ArrowArc(70,70)(10,180,270)
\ArrowLine(150,80)(70,80)
\CArc(70,84)(4,180,270)
\CArc(63,88)(3,0,180)
\Line(60,70)(60,88)
\Line(66,84)(66,88)
\ArrowLine(70,60)(110,60)
\ArrowArc(110,0)(8,90,270)
\ArrowArc(110,20)(8,90,270)
\ArrowArc(110,45)(8,90,270)
\ArrowArcn(150,65)(8,270,90)
\ArrowArcn(150,43)(8,270,90)
\ArrowArcn(150,21)(8,270,90)
\ArrowLine(240,100)(440,100)
\ArrowLine(240,95)(320,95)
\ArrowLine(240,5)(290,5)
\ArrowLine(240,0)(290,0)
\CArc(250,80)(10,0,90)
\Line(240,90)(250,90)
\CArc(270,80)(10,180,270)
\ArrowLine(270,70)(290,70)
\CArc(250,20)(10,-90,0)
\Line(240,10)(250,10)
\CArc(270,20)(10,90,180)
\ArrowLine(270,30)(440,30)
\ArrowArcn(440,83)(8,270,90)
\ArrowArcn(440,62)(8,270,90)
\ArrowArcn(440,41)(8,270,90)
\ArrowArc(290,57)(8,90,270)
\ArrowArc(290,37)(8,90,270)
\ArrowArc(290,18)(8,90,270)
\ArrowLine(300,-15)(350,-15)
\ArrowLine(300,-20)(350,-20)
\CArc(310,0)(10,-90,0)
\Line(300,-10)(310,-10)
\CArc(330,0)(10,90,180)
\ArrowLine(330,10)(390,10)
\ArrowArcn(345,88)(7,270,90)
\ArrowLine(390,81)(345,81)
\ArrowLine(345,95)(440,95)
\ArrowArc(350,0)(8,90,270)
\ArrowArc(350,20)(8,90,270)
\ArrowArc(350,45)(8,90,270)
\ArrowArcn(390,65)(8,270,90)
\ArrowArcn(390,43)(8,270,90)
\ArrowArcn(390,21)(8,270,90)
\CArc(320,85)(10,0,90)
\Line(330,85)(330,70)
\CArc(340,70)(10,180,270)
\ArrowLine(340,60)(350,60)
\Text(100,120)[]{{\large (a)}}
\Text(340,120)[]{{\large (b)}}
\end{picture}
\end{center}
\caption{Two diagrams corresponding to inelastic collisions of a 
nucleon with two nucleons from the target. Diagram (a) is common to 
all models, while diagram (b) is one of the new components considered 
in this paper.} \label{fig:TwoColl}
\end{figure}

This assumption means the inclusion of terms of the topological 
expansion that are missing in the other independent strings 
fragmentation models.  They correspond to diagrams where the 
cylinder (the pomeron) is linked to the colliding hadron through 
a $q^{v}\bar{q}^{s}$ or $\bar{q}^{v}q^{s}$ trajectory as depicted 
in fig. \ref{fig:TwoColl}(b). For the incoming nucleon to interact 
with two 
nucleons of the target it is necessary that it appears in a virtual 
state composed of a baryon-like and a meson-like states. Then, the 
topology of the interaction can be depicted as the superposition of 
a baryon-nucleon and a meson-nucleon collisions. In other models, 
the baryon-like state could only be formed with the three valence 
quarks, 
and only meson-like states of the type $|u\bar{u}>$, $|d\bar{d}>$, 
etc, where considered. I am dropping this restriction only 
justified for high-mass intermediate states, where sea 
quark-antiquark 
pairs are created with large  values of rapidity ($ > 1$) respect 
to their parent hadron. Therefore, in a proton - nucleus collision 
we now have to include, besides $|p^{*} \; \pi^{0*} (\rho, \cdots)>$, 
states like $|\Lambda^{*} \;K^{+*}>$\footnote{ {\em Baryon-like} and 
{\em meson-like}, as well as the asterisks, mean that they are not 
real 
baryons or meson since only the flavour content is considered and no 
attention is paid to other quantum numbers}.

These new components enhance the production of the strange baryons 
in the fragmentation region due to the contribution of strange diquark 
fragmentation. I consider a  theory with  only three flavours: $u$, $d$ 
and $s$, and assume that the $u$ and $d$ contents of the sea are 
equal, while the $s$ content is suppressed and represented by the 
parameter S.
\begin{equation}  \label{eq:DefS}
S = \frac{2\bar{s}}{\bar{u}+\bar{d}} \qquad .
\end{equation} 
SU(3) symmetry corresponds to a value $S=1$. I take $S = 0.5$  that 
is compatible with several data analysis\citeup{ExpHypHE,MartinS,Bazarko}. 
In the rest of the paper, I 
will be using $\gamma$ as the portion of strange sea-quarks over 
the total number of sea-quarks:
\begin{equation} \label{eq:DefGamma}
\gamma = \frac{\bar{s}}{\bar{u}+\bar{d}+\bar{s}} = \frac{S}{S+2}  
\qquad .
\end{equation}
For the given value of $S$, $\gamma =0.2$.

\subsection{Structure Functions} 

The only essential difference between the DPM and the 
QGSM comes from the way they postulate the 
structure functions for sea quarks and antiquarks. Only the leading 
behaviour at $x \rightarrow  0$ is retained in all cases, that has 
been called the {\em retardation} of the hadron fragments. In a 
collision in which there is only one  cut pomeron,  the retardation 
of a quark of flavour $f$ depends on the intercept of the 
corresponding $q^{f}\bar{q}^{f}$ trajectory while for the diquark 
depends on the intercept of the corresponding $qq\bar{q}\bar{q}$ 
exotic trajectory. When dealing with nucleons, for these 
functions we have
\begin{equation}  \label{eq:ValenceSF}
f_{q^v} (x) = x^{-\alpha_R} , \qquad  f_{qq} (x) = x^{\alpha_R (0)-2 
\alpha_N 
(0)} \qquad .
\end{equation}
Where we have made use of the equation, valid in the planar 
approximation, 
$\alpha_{q^{a}q^{b}\bar{q}^{a}\bar{q}^{b}}+\alpha_{q^{c}\bar{q}^{c}} 
= 2\alpha_{q^{a}q^{b}q^{c}}$ and $\alpha_R (0)$  ($\alpha_N (0)$) is 
the zero intercept of the leading Regge (nucleon) trajectory.

When there are more than one cut pomeron, we need quarks and antiquarks from 
the sea to make up enough string ends. In the DPM, it is assumed that 
quark-antiquark pairs are formed as the result of the decay of a 
pomeron which has a retardation given by $1/x$. Following this line 
of reasoning DPM postulates that the retardation of a sea quark or sea 
antiquark is given by
\begin{equation}  \label{eq:SeaDPM}
f_{q^{sea}} ^{ ^{(DPM)}} (x) = \frac{1}{\sqrt{x^2 +(\mu^2 /s) }} \qquad .
\end{equation}
Where $\mu$ is a parameter that sets a mass scale.  QGSM postulates, 
on the other hand, that the retardations of the sea quarks and 
antiquarks have the same origin as that for valence quarks and, 
hence, are the same functions. I adopt this latest view in the 
present work. While there are no practical difference in terms of 
computational results at ISR energies, I find QGSM structure 
functions more coherent with the assumption I have made -- 
sea-valence diquark symmetry. The mechanism 
considered in DPM to postulate the retardation functions, 
must be included in a treatment of the 
problem in which we sum over all possible RFT diagrams. For instance, 
in the case of two inelastic collisions, one term must correspond to 
a coupling between the hadron and the two cut pomerons mediated by 
the exchange of a pomeron (DPM) -- PPP diagram,  another one to the 
same coupling mediated by the exchange of a reggeon (QGSM)-- PRP 
diagram. The corresponding relative weights for this kind of diagrams 
are not known and their computation is an open problem. The choice 
among neglecting one or the other constitutes a model assumption that 
can only be justified by the phenomenology\citeup{StoppingYo}. Yet, 
the eikonal approximation to RFT points that the PPP diagrams 
constitute the main contribution to high-mass intermediate states and 
eventually lead to the point-like quarks structure functions, while 
the PRP type of diagrams are the dominant terms for low-mass 
intermediate 
states\citeup{KaidalovPC}. This idea allowed to understand the 
data 
from HERA on small-x behaviour of the structure 
functions\citeup{CKMT}.

The PPP coupling does not allow the  
inclusion of diagrams of type \ref{fig:TwoColl}(b). The 
extra sea quark-antiquark pairs, in 
collisions with more than one cut pomeron, are the result of the 
decay of an object with vacuum quantum numbers. 
Therefore, DPM only allows diagrams of type \ref{fig:TwoColl}(a). 
This also explains why only valence diquarks contribute to the 
final state in DIS and hard hadron-hadron collisions. Then, 
as far as we are interested in the low-$p_{\bot}$ regime, it seems 
most reasonable to adhere to QGSM view.

The s-quark, $(us)$, $(ds)$ and $(ss)$ retardations are 
given by the following functions, where $\alpha_{\varphi}$ is 
the intercept of the $\varphi$ trajectory
 and $\Delta \alpha = \alpha_{R} (0) - \alpha_{\varphi} (0)$.

\begin{eqnarray}
f_{s} (x)  & = & x^{\alpha_{\varphi}} \qquad ,   \nonumber  \\
f_{us} (x) & = & f_{ds} (x) = x^{\alpha_R (0)-2\alpha_N (0)+ \Delta 
\alpha}   \qquad ,
\label{eq:StrRetu}  \\
f_{ss} (x) & = & x^{\alpha_R (0)-2\alpha_N (0)+ 2 \Delta \alpha}  \qquad .
\nonumber 
\end{eqnarray}
The corresponding functions for the $uu$, $ud$, and $dd$ diquarks, as well 
as for the $u$ and $d$ quarks are given in equation \ref{eq:ValenceSF}.

The momentum-fraction distribution function in a nucleon, in the case 
of $N$ cut pomerons, is given by the following expression:
\begin{eqnarray}        
\varrho^{N}(x_{i_{1}},x_{i_{2}},\ldots,x_{i_{N}},x_{i_{d}},x_{\bar{i}_{2}},    
\ldots,
    x_{\bar{i}_{N}})   =  
C^{N}_{i_{1},i_{2},\ldots,i_{N},i_{d},\bar{i}_{2},
    \ldots,\bar{i}_{N}}  \nonumber \\
    \times \; f_{i_{1}} (x_{i_{1}}) \; f_{i_{2}} (x_{i_{2}}) \; 
\ldots \; 
    f_{i_{N}} 
    (x_{i_{N}})  \; f_{i_{d}} (x_{i_{d}}) \; f_{\bar{i}_{2}} 
    (x_{\bar{i}_{2}}) \; \ldots \; 
    f_{\bar{i}_{N}} (x_{\bar{i}_{N}}) \label{eq:TotStrFunc} \\
    \times \; \delta 
(x_{i_{1}}+x_{i_{2}}+\cdots+x_{i_{N}}+x_{i_{d}}+x_{\bar{i}_{2}}+\cdots+
x_{\bar{i}_{N}} - 1)  \qquad . \nonumber 
\end{eqnarray}

This function is perfectly defined if we consider that 
$f_{\bar{i}} (x) = f_{i} (x)$. The indexes $i_{j}=\{ u,s,d \}$ and 
$i_{\bar{j}} = \{  \bar{u},\bar{d},\bar{s} \} $, $i_{d}= \{ 
(uu),(ud),(dd),(us),(ds),(ss) \}$. 
Since we are considering nucleons as the colliding particles, 
the diquark can be of type $(ss)$ only if $N \geq 3$ and of types 
$(us)$ or $(ds)$ if $N \geq 2$; we find the s-quark 
retardation only when  $N \geq 2$.

The single-parton structure functions are obtained integrating over 
all other momentum fractions:
\begin{eqnarray}
\rho_{u,d}^{(N)}(x) &=& C_{0}^{q,N} x^{-\alpha_R (0)}    
(1-x)^{\alpha_R (0)-2\alpha_N (0)+(N-1)2(1-\alpha_R (0))} \nonumber \\
& & \times \; [1 - \gamma + \gamma (1 -  x) ^{2 \Delta \alpha} ]^{N-1}
    \nonumber \\
&=& C_{0}^{q,N} x^{-\frac{1}{2}}    (1-x)^N (1- 
\gamma x)^{N-1}, \qquad N \geq 1   \qquad .   \nonumber \\
& &     \nonumber \\
\rho_{s}^{(N)} (x)  &=& C_{1}^{q,N}   x^{-\alpha_{\varphi} (0)}
(1-x)^{\alpha_R (0)-2\alpha_N (0)+2(N-2)(1-\alpha_R (0))+
2-\alpha_R (0)-\alpha_{\varphi} (0)}  \nonumber \\
& & \times \; [1 - \gamma + \gamma (1 -  x) ^{2 \Delta \alpha} ]^{N-2}
  \nonumber \\
&=& C_{1}^{q,N} (1-x)^{N+\frac{1}{2}} (1- \gamma x)^{N-2}, \qquad N 
\geq 2   \qquad . 
\nonumber \\
& &     \nonumber \\
\rho_{uu,ud,dd}^{(N)} (x) &=& C_{0}^{qq,N} x^{\alpha_R (0)-2\alpha_N 
(0)}
(1-x)^{-\alpha_R (0) + 2 (N-1) (1-\alpha_R (0))}  \nonumber \\
& & \times \; [1 - \gamma + \gamma (1 -  x) ^{2 \Delta \alpha} ]^{N-1}
  \nonumber \\
&=& C_{0}^{qq,N} x(1-x)^{N-\frac{3}{2}} (1- \gamma x)^{N-1}, \qquad N 
\geq 1    \qquad . 
\label{eq:StructFunc} \\
& &     \nonumber \\
\rho_{us,ds}^{(N)} (x) &=& C_{1}^{qq,N} x^{\alpha_R (0)-2\alpha_N 
(0)+ 
\Delta \alpha} 
(1-x)^{-\alpha_{\varphi} (0)+2(N- 1)(1-\alpha_R (0))}  \nonumber \\
& & \times \; [1 - \gamma + \gamma (1 -  x) ^{2 \Delta \alpha} ]^{N-2}
  \nonumber \\
&=& C_{1}^{qq,N} x^{\frac{3}{2}} 
(1-x)^{N- 1} (1- \gamma x)^{N-2}, 
\qquad N \geq 2  \qquad .  \nonumber \\
& &     \nonumber \\
\rho_{ss}^{(N)} (x) &=& C_{2}^{qq,N}   x^{\alpha_R (0)-2\alpha_N (0)+ 
2 
\Delta \alpha} 
(1-x)^{-\alpha_R (0) + 2 (N-2) (1-\alpha_R (0))+2(1-\alpha_{\varphi} 
(0))}  \nonumber \\
& & \times \; [1 - \gamma + \gamma (1 -  x) ^{2 \Delta \alpha} ]^{N-3}
  \nonumber \\
&=& C_{2}^{qq,N} x^2                (1-x)^{N- 
\frac{1}{2}} (1- \gamma x)^{N-3}, \qquad N \geq 3  \qquad .  \nonumber
\end{eqnarray}

The numerical values for the constants are  $\alpha_R (0) = 0.5$, 
$\Delta \alpha =0.5$ and $\alpha_{N}=  -0.25$. The value of $\alpha_{N}$ has 
been chosen within the theoretically allowed range to obtain the best 
agreement with the proton-proton data.

The factors $[1 - \gamma + \gamma (1 -  x) ^{2 \Delta \alpha} 
]^{N-J}$ in equation (\ref{eq:StructFunc}) have their origin in the 
fact that the $s$ quarks take a larger portion of the nucleon momentum that 
the $u$ or $d$ quarks. It has a little effect on the structure functions 
when $N$ is small. When $N$ is large, these factors make the 
functions steeper at $x \rightarrow 0$.

\subsection{Fragmentation Functions} \label{subsec:SF}

The fragmentation functions are derived from Regge 
arguments\citeup{KaidalovFragmentation,KaidalovPiskunova}. The ones 
needed for the present work are listed in the appendix. I have corrected 
some missprints in ref. \cite{KaidalovPiskunova} and use 
consistently the value of $\alpha_{N}$ given above\footnote{The 
treatment of the fragmentation functions in ref. 
\cite{CapellaMerinoStrange} 
is inconsistent since two different values for $\alpha_{N}$, 
$-0.25$ and $-0.5$, were used simultaneously. This second 
value was also used to compute the structure functions.}.
The constant $\lambda = 2\alpha' \bar{p}_{\bot}^{2} \simeq 0.5 $, 
where $\alpha'$ is the slope or the Regge trajectory, has been used 
in the expressions.

All these functions are proportional to one of the constants $a_{p}$, 
$a_{\Lambda}$, $a_{\bar{p}}$ and $a_{\bar{\Lambda}}$, which represent 
the universal rapidity densities of protons, $\Lambda$ hyperons, antiprotons 
and anti-$\Lambda$ hyperons respectively, around the 
centre-of-mass of the string. The fragmentation of a diquark into a 
given baryon contains two terms, one corresponds to the creation 
of a hadron that contains the baryon number of the diquark: 
primary or leading fragmentation (subindex 1), the  other  
comes from baryon-antibaryon formation in the string: secondary 
fragmentation (subindex 2). Hence, the latest is proportional 
to the corresponding antibaryon density.
Functions such as $D_{1,uu}^{p}(z)$, include a factor 
function linear with $z$. This is introduced to take into account the 
excess in the production of baryons containing the whole diquark 
over those that contains only one of the quarks of the diquark. 
Taking into account the constants that define these factors, 
$c_{0}$, $c_{1}$ and $c_{2}$, we have seven parameters with only 
three of them independent as a consequence of the baryon number
sum rule. \footnote{In ref. 
\cite{CapellaMerinoStrange} the constraint imposed by baryon number 
conservation was overlooked and different values for the baryon 
densities in the strings were used.}.
I have assumed that this sum rule saturates with only 
protons, neutros and $\Lambda$'s as if they were the only
stable baryons under strong interactions, being all the
strong decays taken into account by the Regge treatment
of the problem. In fact there are other hyperons that decay through
electroweak processes into $\Lambda$'s and nucleons which have some
contribution to the final rapidity spectra. We can easily 
include in the treatment these
hyperons assuming that the corresponding fragmentation functions are
proportional to the ones for the $\Lambda$ production. This is strictly 
true for the cases of $\Sigma^{+}$,$\Sigma^{0}$ and  $\Sigma^{-}$ while 
it is only approximate for the case of $\Xi$ hyperons which 
provide a much smaller relative
contribution. I take the contribution from this decays 
making use of the available phenomenological information. The total 
number of hyperons (anti-hyperons) is taken to be $1.6$ the number of $\Lambda$'s
($\bar{\Lambda}$). $30\%$ of the other strange hypeons (anti-hyperons) 
decay into $\Lambda$ ($\bar{\Lambda}$)\citeup{NA35Lambda,NA49proton}. 
Hence, with all these considerations
taken, we may interpret $a_{\Lambda}$ ($a_{\bar{\Lambda}}$) as the hyperon 
(anti-hyperon) density in the middle of the string some of which will decay 
into  $\Lambda$ ($\bar{\Lambda}$) and others into protons (antiprotons) and 
neutrons (antineutrons) in equal proportions. The corrections introduced 
by these consideration are not significant, in the case of $\Lambda$ 
distributions,  given the accuracy of the experimental data as will 
be seen latter in the following sections.

The functional form of the diquark fragmentation functions for diquarks with 
strangeness $-1$ and $-2$ were obtained applying the results of 
ref. \cite{KaidalovFragmentation}. I have assumed that 
both strange and non-strange 
quarks are dynamically symmetric, hence the factor $(1+\sqrt{z})/2$. The 
fragmentation of a diquark into a baryon that do not 
contain any quark from the diquark is taken to be $0$. This may be 
controversial since it has been suggested\citeup{KaidalovPC} that 
an excess in the production of $\Omega^{-}$ over $\Omega^{+}$, in 
non-strange hadrons collisions, contradicts this assumption. As a 
matter of fact, preliminary data show 
a very small asymmetry in  $\pi$-nucleus interactions at $500 
\mbox{GeV/c}$\citeup{Solano}. Nevertheless, although the data are 
still inconclusive, this effect, if small as it seems, may be 
understood in the framework of the model described in this paper: 
multiple collision raises the probability of having contributions 
to the final state coming from the fragmentation of strange diquarks 
that results in a net production of $\Omega^{-}$. 

\subsection{Rapidity Distributions}

We concentrate in the study of collisions between identical nuclei. 
The atomic weight and the atomic number are $A$ and $Z$ respectively. 
The rapidity distribution for the production of a hadron $h$ is given 
by:
\begin{eqnarray} 
\frac{dN_{AB}^{(h)}}{dy} & = & \frac{1}{\sigma_{AB}} 
\sum_{(n_{A},n_{B},n,\mu_{A},\mu_{B})}
\sigma^{n_{A},n_{B},n,\mu_{A},\mu_{B}} \{
\theta(n_{B}-n_{A}) [ n_{A} (N^{qq^{A}-q^{B}}_{\mu_{A}\mu_{B}} (y) +
N^{q^{A}-qq^{B}}_{\mu_{A}\mu_{B}} (y)) +  \nonumber  \\
& &(n_{B}-n_{A})(N^{\bar{q}^{A}-q^{B}}_{\mu_{A}\mu_{B}} (y) +
N^{q^{A}-qq^{B}}_{\mu_{A}\mu_{B}} (y))  +
(n-n_{B}) (N^{q^{A}-\bar{q}^{B}}_{\mu_{A}\mu_{B}} (y) +
N^{\bar{q}^{A}-q^{B}}_{\mu_{A}\mu_{B}} (y)) ]  \label{eq:YDist1} \\
& &+ \hbox{Symmetric terms}(n_{A} \leftrightarrow n_{B})
\}  \qquad .  \nonumber
\end{eqnarray}
In writing this expression, I neglect kinematical correlations 
among different strings, $N^{a-b}_{\mu_{A}\mu_{B}}$. Here, 
$\sigma^{n_{A},n_{B},n,\mu_{A},\mu_{B}}$ is the cross-section 
for $AB$ collisions where there are $n$ inelastic nucleon-nucleon 
collisions involving $n_{A}$ nucleons of $A$ and $n_{B}$ nucleons 
of $B$, in which at least one of the participating nucleons of 
$A$ ($B$) suffers $\mu_{A}$ ($\mu_{B}$) inelastic collisions.

I restrict the discussion to the case $A=B$. We can approximate 
equation \ref{eq:YDist1}, with a good degree of accuracy using 
average values for $n_{A}$, $n_{B}$ and $n$.
\begin{equation} \label{eq:YDist2}
\frac{dN_{AB}^{(h)}}{dy}= \bar{n}_{A} 
(N^{qq^{A}-q^{B}}_{\bar{\mu}_{A} \bar{\mu}_{A}} (y) +
N^{q^{A}-qq^{B}}_{\bar{\mu}_{A} \bar{\mu}_{A}} (y) ) +
 (\bar{n}-\bar{n}_{B}) 
(N^{q^{A}-\bar{q}^{B}}_{\bar{\mu}_{A} \bar{\mu}_{A}} (y)   +
N^{\bar{q}^{A}-q^{B}}_{\bar{\mu}_{A} \bar{\mu}_{A}}  (y)  )
\qquad .
\end{equation}

Equation (\ref{eq:YDist2}) is obtained by neglecting the dependence 
of the string density $N$ on the $\mu$'s, which are replaced by 
their average values $\bar{\mu}$
\footnote{In practical computations a slightly more accurate 
expression has been used in which the average of densities 
over $\mu_{B}=\mu_{A}$ were taken using the weights 
$\sigma_{NA}^{\mu_{A}}/\sigma_{NA}$. 
These weights are obtained from ref. \cite{LeadingBaryonYo}. 
No significant numerical difference were observed compared to 
the results obtained using $\bar{\mu}_{A}\simeq\bar{n}/\bar{n}_{A}$.}.
\begin{eqnarray} 
N^{a-b}_{\mu_{1},\mu_{2}} (y) &=& \sum_{i,j} 
w_{N}^{i} w_{N}^{j}
\int_{0}^{1} 
\int_{0} ^{1} dx_{1} dx_{2} 
\varrho^{a_{i}}_{\mu_{1}} (x_{1}) 
\varrho^{b_{j}}_{\mu_{2}} (x_{2}) 
\frac{dN^{a_{i}-b_{j}}}{dy} (y-\Delta,s_{h}^{a_{i}-b_{j}}) 
\label{eq:StringDens} \\
& &\theta(sx_{1}x_{2}-s_{h}^{a_{i}-b_{j}}) \qquad . \nonumber
\end{eqnarray}

Here $a$ and $b$ refer to quark or diquark.
A sum is performed over all possible partons flavours 
$i$ and $j$. The probability weights $w^{i}$ and $w^{j}$ 
are trivially computed from the model assumptions. They depend on 
$\gamma$, $A$ and $Z$. A mass threshold for each chain 
is enforced by the $\theta$-function. Please note that 
this mass threshold is not always equal to the mass 
of hadron $h$ since the flavour structure of the string 
many times forces to have other hadrons besides $h$ in 
the final state . $\Delta$ is the center-of-mass rapidity 
of the string.

\begin{equation} \label{eq:NGes}
\frac{dN^{a_{i}-b_{j}}}{dy} (y-\Delta,s_{h}^{a_{i}-b_{j}}) = 
\left\{ \begin{array}{ll}
       G_{a_{i}}^{h}(y) & \; \mbox{if}  \; y\geq \Delta \qquad ,  \\
       G_{b_{j}}^{h}(y) & \; \mbox{if}  \; y\leq \Delta \qquad , 
       \end{array}
\right.
\end{equation}

\noindent where $G(y)$ are the fragmentation functions:

\begin{equation} \label{eq:Ges}
 \begin{array}{ll}
G_{a_{i}}^{h}(y) &  = Z_{+}D^{h}_{a_{i}}(Z_{+}) \qquad ,  \\
G_{b_{j}}^{h}(y) &  = Z_{-}D^{h}_{b_{j}}(Z_{-}) \qquad .
       \end{array}
\end{equation}

\noindent Here $Z_{+} = \exp[(y-\Delta)-y_{MAX}]$ and 
$Z_{-} = \exp[-(y-\Delta)-y_{MAX}]$, $y_{MAX}$ being the 
maximum rapidity that hadron $h$ can have in the string 
for fixed values of momentum fractions $x_{1}$ and $x_{2}$.

\section{Proton-proton collisions}

Since I have not considered sum rules others than the ones 
due to baryon conservation, the model leaves three free 
parameters that I have fixed in order to obtain the best 
fits to proton-proton data. At the ISR energies  the 
structure of these collisions is very simple as we only 
need to consider one pomeron exchange\citeup{DPMPhysRep}. 
This makes the choice of this energy range particularly 
convenient as we avoid having to take into account the probability 
weights for the different re-scattering configurations within 
the same nucleon-nucleon collision. The 
only differences between the results of the present model and 
those from DPM and QGSM, at this stage, come from the choice 
of $\alpha_{N}$, that was usually taken in the past to be 
near $-0.5$.

\begin{figure}[p]
\epsfbox[0 370 300 700]{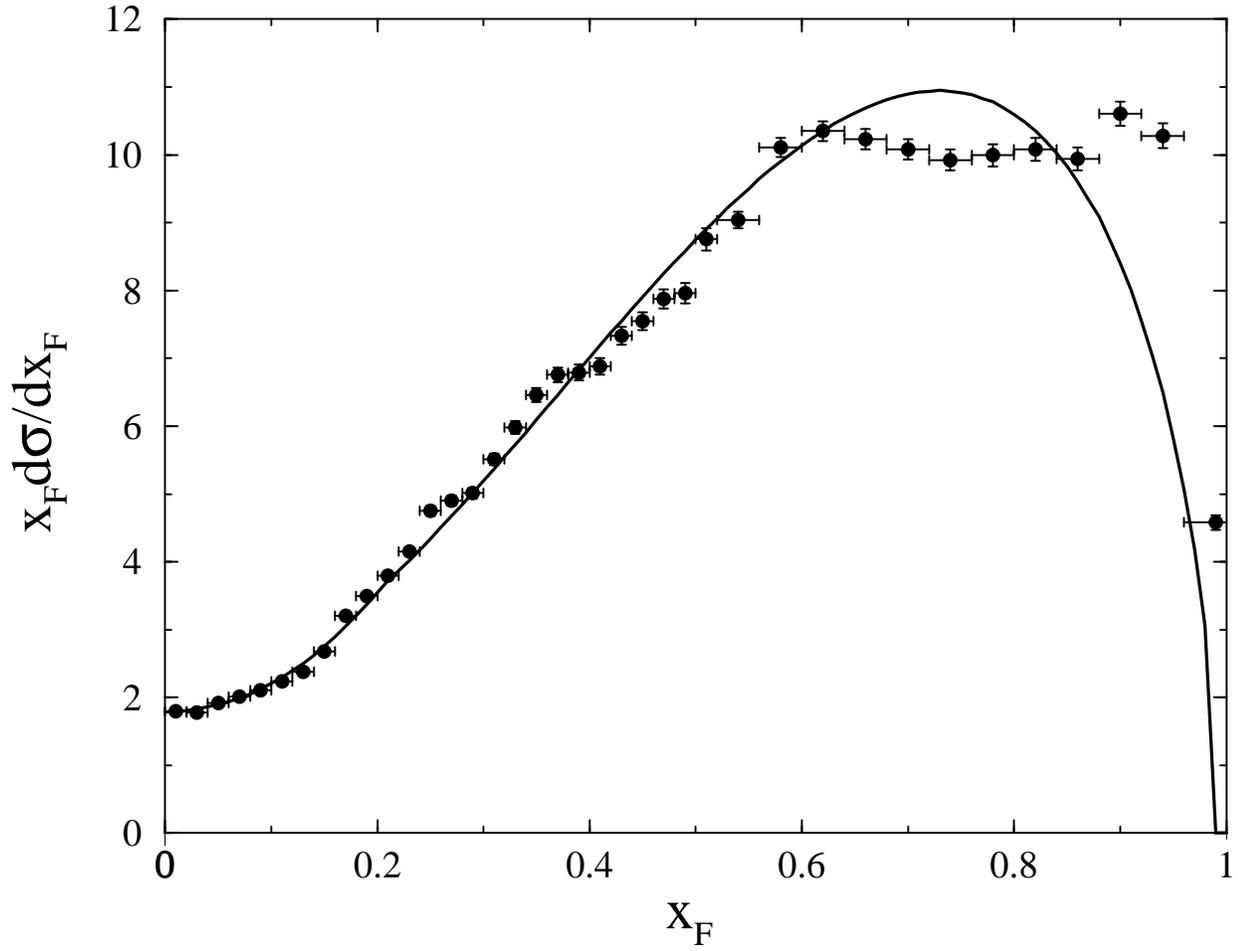}
\caption{Fit to the Feynman $x$ distribution 
of protons in the reaction $pp \rightarrow pX$. Data correspond 
to $p_{\mbox{lab}} = 400\mbox{ GeV}$ and were taken from  
ref. \cite{ExpPPPPB}.} \label{fig:Graph1} 
\end{figure}

\begin{figure}[p] 
\epsfbox[0 370 300 700]{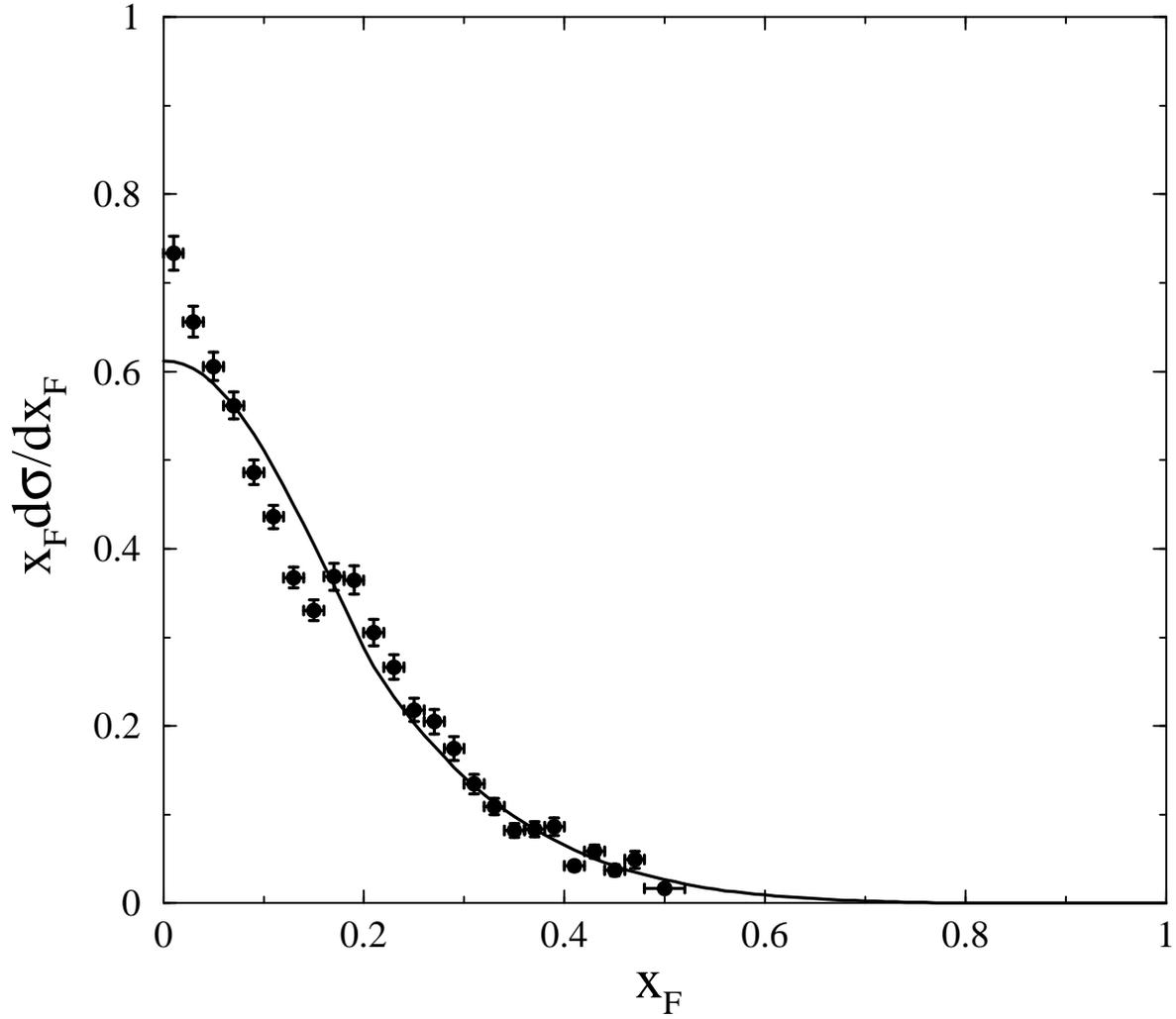}
\caption{Fit to the Feynman $x$ distribution 
of antiprotons produced in the reaction $pp \rightarrow \bar{p}X$. Data 
correspond to $p_{\mbox{lab}} = 400\mbox{ GeV}$ and were 
taken from  ref. \cite{ExpPPPPB}.} \label{fig:Graph2}  
\end{figure}

Figure \ref{fig:Graph1} represents the Feynman $x$ distribution 
of protons produced in the reaction $pp\rightarrow pX$. The fit 
obtained agrees very well with the experimental data taken from 
ref. \cite{ExpPPPPB} particularly in the region 
$x_{F} < 0.75$. It has been pointed out that experimental 
points above that cut are not very reliable 
(see  \cite{ExpPPPPB}) due to detector inefficiencies. 
The fit in the region 
$x \rightarrow 0$ is very sensitive to the value of  $\alpha_{N}$. 
For $\alpha_{N}=-0.5$ the results of the fit are much poorer that 
the one shown, which corresponds to the value of $\alpha_{N}=-0.25$ 
consistently used throughout this paper. The value of $a_{p}$ fixes 
the values of the remaining parameters by the use of the sum rules.
Figure \ref{fig:Graph2} 
shows the fit to $pp \rightarrow \bar{p} X$. It is also worth 
mentioning that the same value for 
$a_{\bar{p}}$ was independently obtained from both fits. Please 
note that the computation performed do not include diffraction 
and, consequently, the results obtained do not show the characteristic 
diffraction pick at large-$x$.

\input epsf

\begin{figure}[p] 
\epsfbox[0 390 300 700]{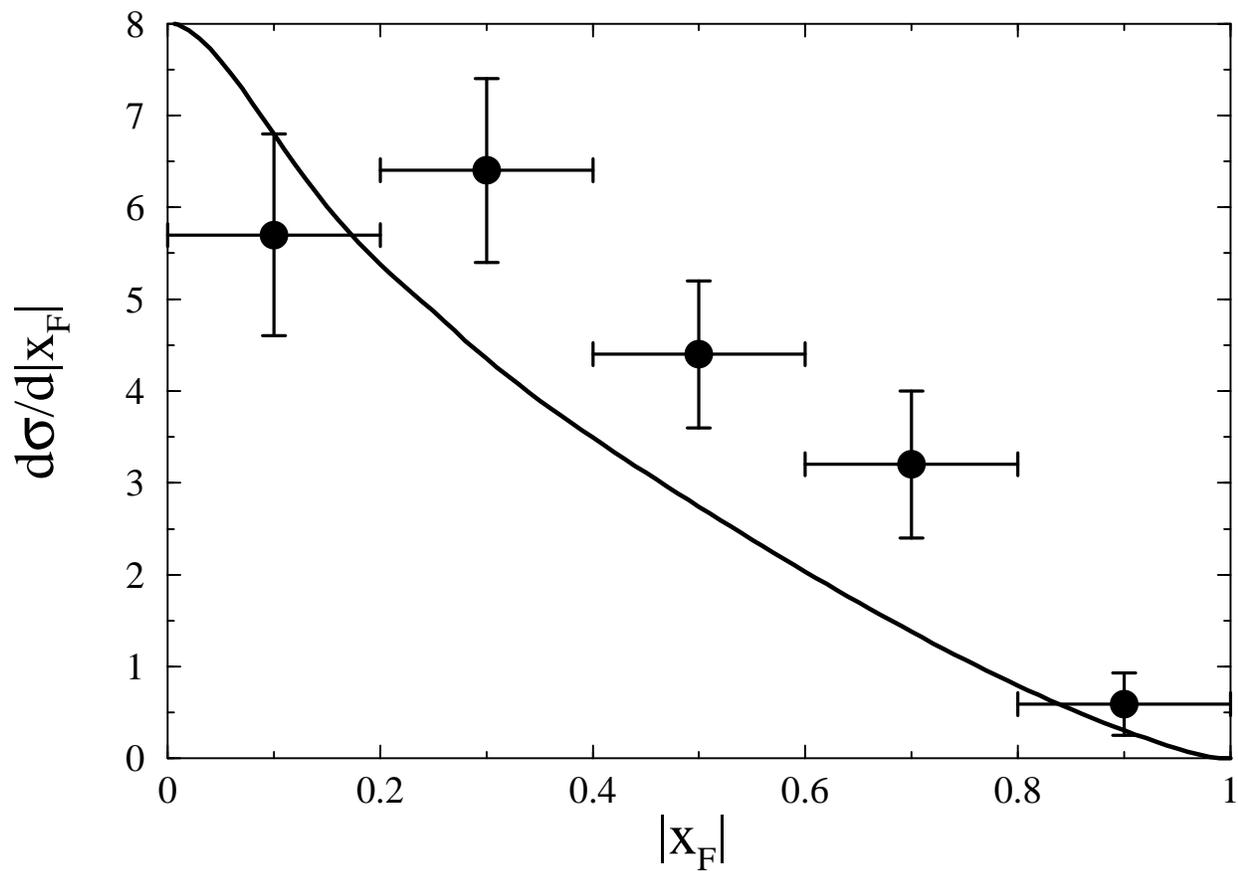}
\caption{Fit to the Feynman $x$ differential cross section
 $d\sigma/dx$ (given in mb)  
for the inclusive production of $\Lambda$ in 
the reaction $pp \rightarrow \Lambda X$.  Data 
correspond to $p_{\mbox{lab}} = 405\mbox{ GeV}$ and were 
taken from ref. \cite{ExpPPLLB}.} \label{fig:Graph3}
\end{figure}

\begin{figure}[p]
\epsfbox[0 370 300 670]{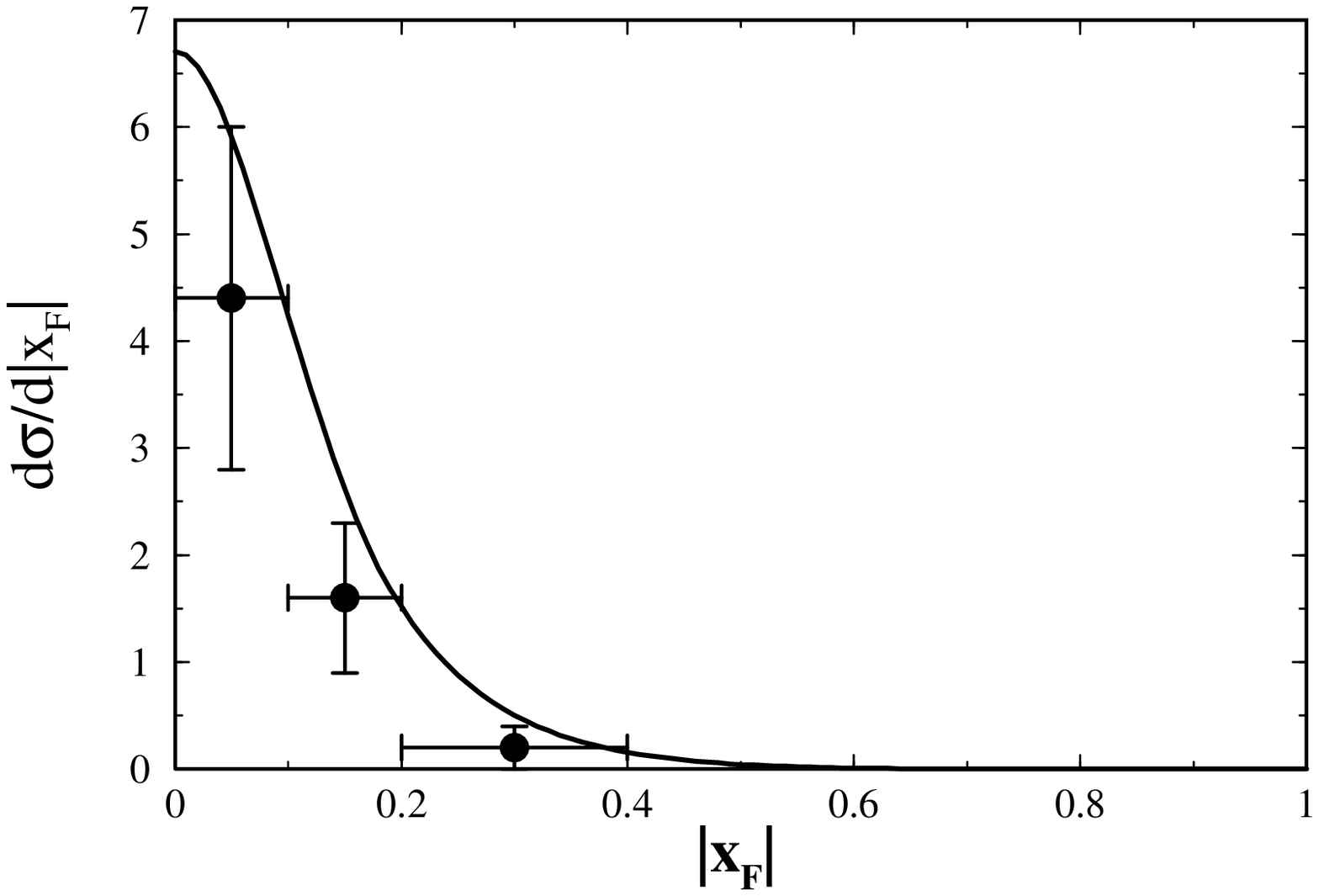}
\caption {Fit to the Feynman $x$ differential cross section
$d\sigma/dx$ (given in mb)  
for the inclusive production of $\bar{\Lambda}$ in 
the reaction $pp \rightarrow \bar{\Lambda} X$. Data 
correspond to $p_{\mbox{lab}} = 405\mbox{ GeV}$ and 
were taken from ref. \cite{ExpPPLLB}.} \label{fig:Graph4} 
\end{figure}

To obtain de value of $a_{\bar{\Lambda}}$, I used the data from 
ref. \cite{ExpPPLLB} that were taken at a similar centre-of-mass 
energy. The reactions considered 
are $pp\rightarrow \Lambda X$ and $pp\rightarrow \bar{\Lambda} X$. 
We show the best fits in figures \ref{fig:Graph3} and  
\ref{fig:Graph4} (please mind that in this two fits 
only $a_{\bar{\Lambda}}$ is free, since  $a_{\Lambda}$ 
is given by the sum rules and the value of  $a_{p}$)
The large error bars in the experimental data do not allow for 
a very accurate determination of $a_{\bar{\Lambda}}$. Corrections from
contributions of electroweak decays of heavier hyperons, as discussed 
in subsection \ref{subsec:SF}, have also been considered. I have 
taken the value that best fits the data and also that is compatible 
with the mass suppression factor of introduced in 
\cite{KaidalovFragmentation}.

The numerical values obtained are:  
$a_{p} = 0.667$, 
$a_{\Lambda} = 0.282$, $a_{\bar{p}} = 0.077$,  
$a_{\bar{\Lambda}} = 0.057$, $c_{0}=1.430$, $c_{1}=6.076 $ 
and $c_{2}= 67.560$. 

\section{Nucleus-Nucleus collisions}

At this point we have a model for nucleus-nucleus collisions 
which contains no free parameters. 

\begin{figure}[p]
\epsfbox[0 370 300 700]{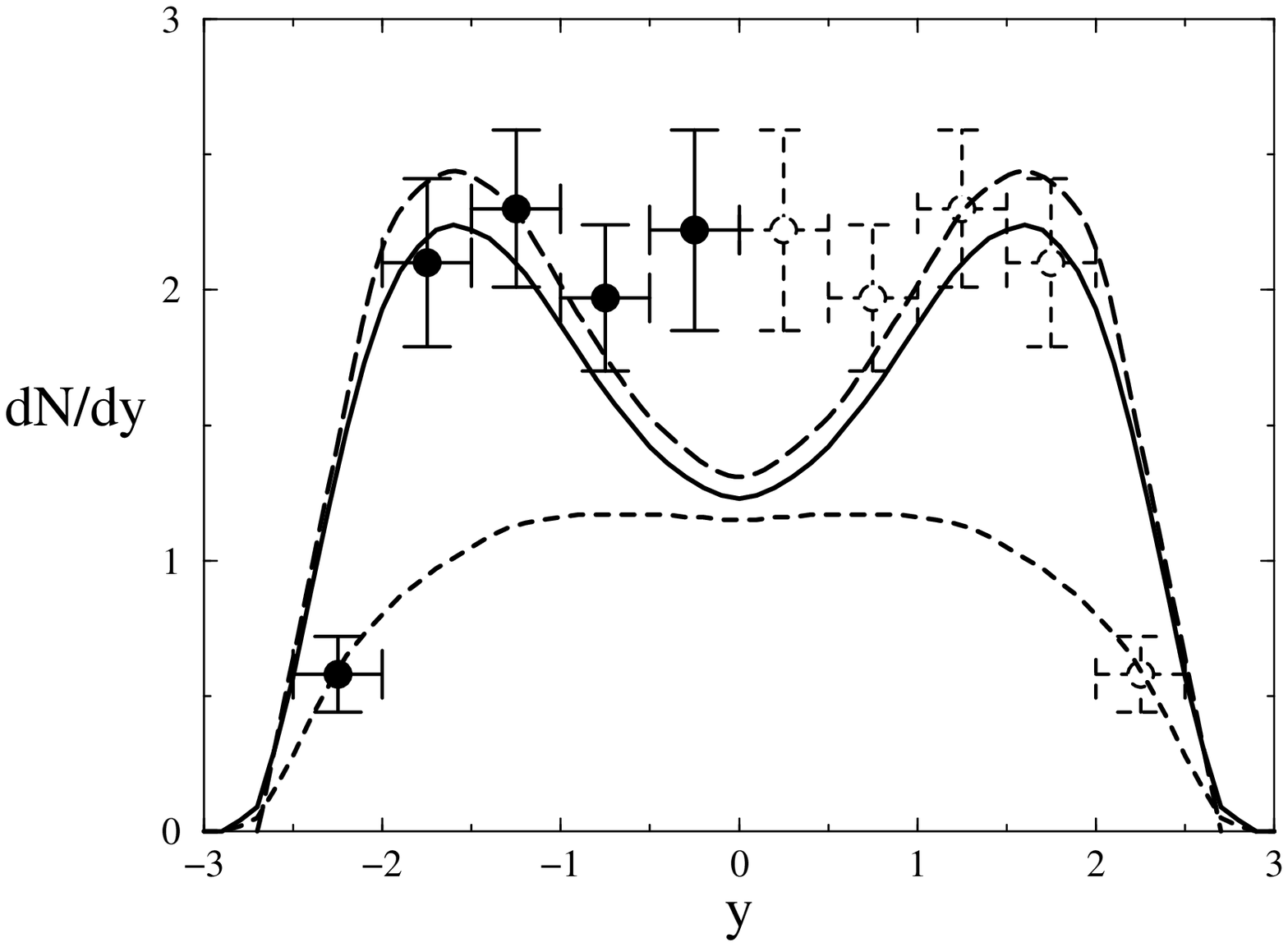}
\caption{Rapidity distribution of $\Lambda$'s produced in central 
sulphur-sulphur collisions represented in the nucleon-nucleon 
centre-of -mass reference frame. The solid line correspond 
to the results of this model while the short-dashed line are 
the results obtained when the new components are not included. 
The long dashed line represent the total strange-hyperons spectrum.
The data points are from ref.\cite{NA35Lambda} 
and have been reflected onto positive rapidities.} \label{fig:Graph5} 
\end{figure}

\begin{figure}[p]
\epsfbox[0 370 300 700]{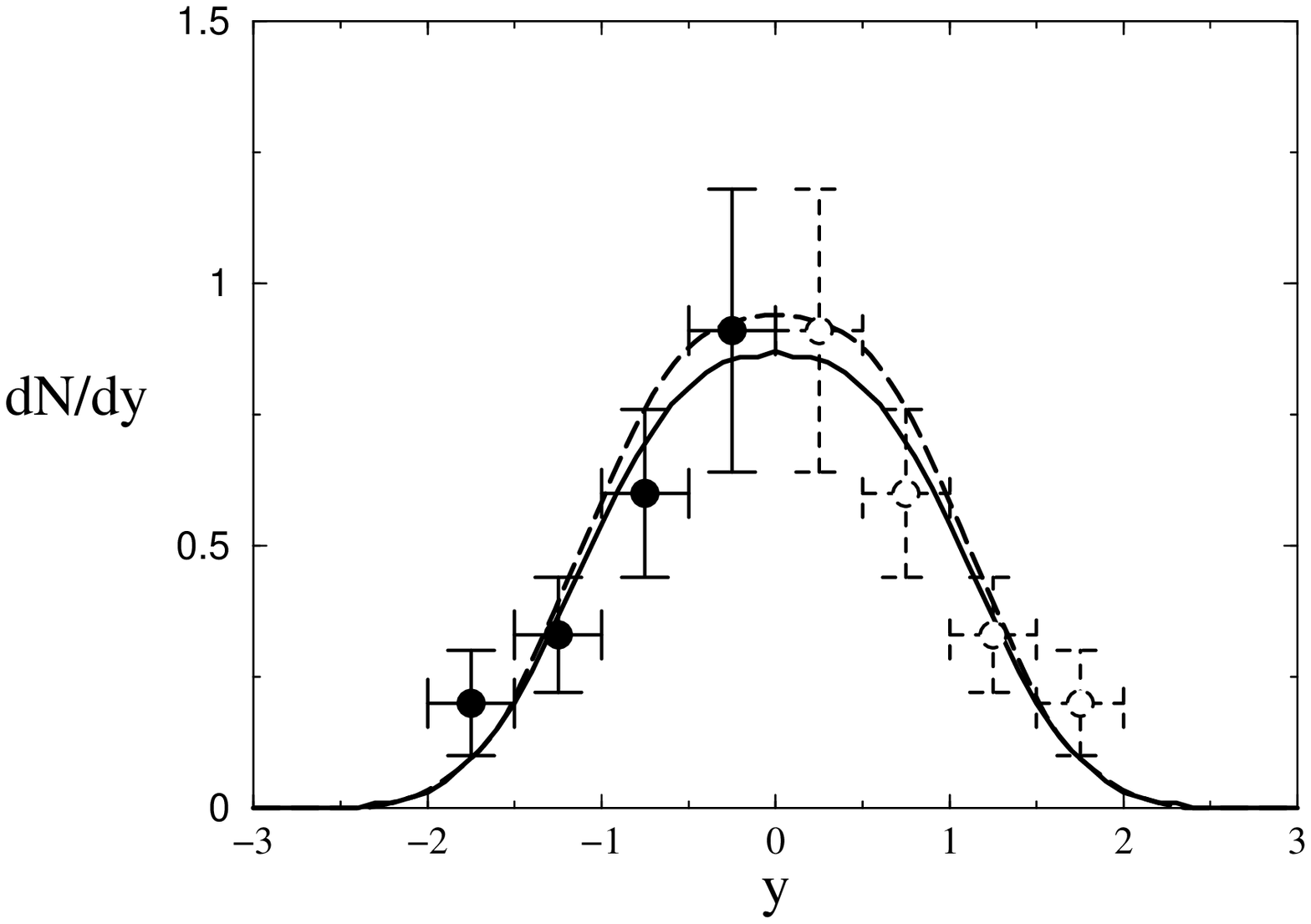}
\caption{Rapidity distribution of $\bar{\Lambda}$'s produced in central 
sulphur-sulphur collisions represented in the nucleon-nucleon 
centre-of-mass reference frame (solid line) and the total strange 
hyperons spectrum (dashed line) . The data points are 
from ref. \cite{NA35Lambda} and have been reflected onto 
positive rapidities.} \label{fig:Graph6} 
\end{figure}

The result for the rapidity distribution of $\Lambda$ hyperons 
produced in central sulfur-sulfur collisions is compared  with 
the experimental data from the CERN-NA35\citeup{NA35Lambda} 
Collaboration in figure \ref{fig:Graph5}. The short-dashed line 
corresponds to the results of the model without the sea-valence 
diquark symmetry introduced in this paper. As we can see, both 
models give similar results at $y_{CM}=0$, 
while there is a remarkable difference in the fragmentation 
region where only the present model  reproduces the data. The 
data points are not corrected for  particles originated 
from electromagnetic $\Sigma^{0}$ ($\bar{\Sigma}^{0}$) decays 
and from weak $\Xi$ ($\bar{\Xi}$) decay. These decays have been
taken into accout as described in \ref{subsec:SF}. The uncorrected 
results are showed  with a long-dashed line and represent the total
strange hyperon distribution. In computing the correction an 
effective recoil, due to the decay of the resonances 
of 0.2 units in rapidity was introduced which has none or very 
little effect on the results. 
Figure  \ref{fig:Graph6} shows the agreement between the 
$\bar{\Lambda}$ distribution and the data points from the 
same experiment. Again, the dashed line corresponds to the
uncorrected hyperon distribution.
The only disagreement is with the point 
near $y_{CM}=0$ in the $\Lambda$ distribution. 


\begin{figure}[p] 
\epsfbox[0 370 300 750]{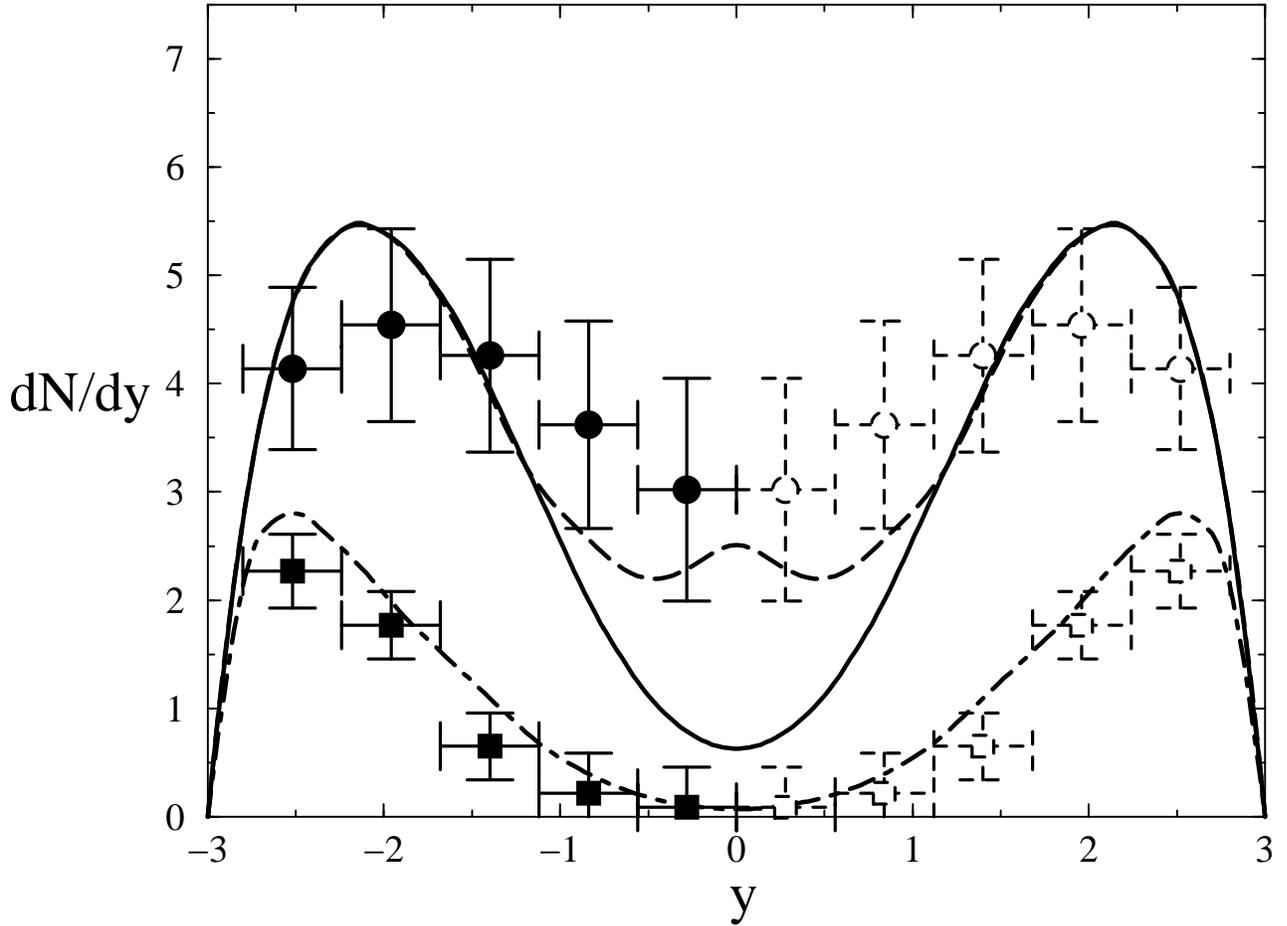}
\caption{$p-\bar{p}$ rapidity distribution in 
sulphur-sulphur collisions represented in the nucleon-nucleon 
centre-of-mass reference frame. The dashed-dotted line 
corresponds to primary 
protons produced in peripheral collisions; the solid  
line are the results for the proton minus antiproton 
distribution in central collisions.The data points are from 
ref. \cite{NA35Proton} and have been reflected onto positive 
rapidities. The dashed line represents the correction for the 
excess of $K^{+}-K{-}$. See text for a more 
detailed explanation.} \label{fig:Graph7}
\end{figure}

Figure \ref{fig:Graph7} shows the proton minus antiproton
distribution  in central (solid line) and peripheral 
(dashed-dotted line) sulfur-sulfur collisions. 
The data points were also taken by the CERN-NA35 
collaboration\citeup{NA35Proton}. For central collisions, 
the agreement in the fragmentation region of the spectrum 
is very good while the data show an overpopulation of 
protons around $y_{CM}=0$. The data were extracted from
the measured charge asymmetry of the final state applying 
model dependent corrections. The cancellations of  
$p-\bar{p}$ and $\pi^{+}-\pi^{-}$ have been assumed
due to the isospin symmetry of the system. The results showed 
here do not make use of the first assumption above. A small difference, 
not significant in magnitude, is obtained in the central 
rapidity region when we compute the leading proton distribution 
as if the  $p-\bar{p}$ cancellation was exact, leaving the 
integrated correction equal to zero as it should. From 
quark flavour counting the model predicts  an overal 
$\pi^{+}-\pi^{-}$ cancellation and an negligible change on 
this result due to any local effect.
Corrections based on a Monte Carlo simulation (Fritiof)  to the data 
have also been applied to estimate the $K^{+}-K^{-}$ contribution 
to the charge asymmetry. The model presented here predicts a much 
higher production of $\Lambda$, in agreement with the data, than
Fritiof whose results are very close to those of this model without
the new components (dashed line fig. \ref{fig:Graph5}). For each 
extra produced lambda we are left with a $\bar{s}$ quark at the
end of one string that will appear in the final state as a $K^{+}$ or 
a  $K^{0}$ in equal proportions. The obvious conclusion is that
this model predicts a higher $K^{+}-K^{-}$ asymmetry concentrated
in the central rapidity region and whose total value is equal to 
one half the excess of  $\Lambda$ predicted by this model over
the predictions of Fritiof. I have added a Gaussian correction to
the $p-\bar{p}$ distribution that accounts for the missed point
in the central region (dashed line). Given the long error bars, 
the precise shape of this correction do not matter significantly.
It is interesting to notice that another correction to the 
 $\Lambda$ distribution might be necessary due to the effect
on the $\Lambda \rightarrow p \pi^{-}$ background coming from 
$K^{0} \rightarrow \pi^{+} \pi^{-}$. Although, there
is not enough information available to assume that this background
has not been correctly subtracted in the experimental analysis.

\begin{figure}[p]
\epsfbox[0 370 300 720]{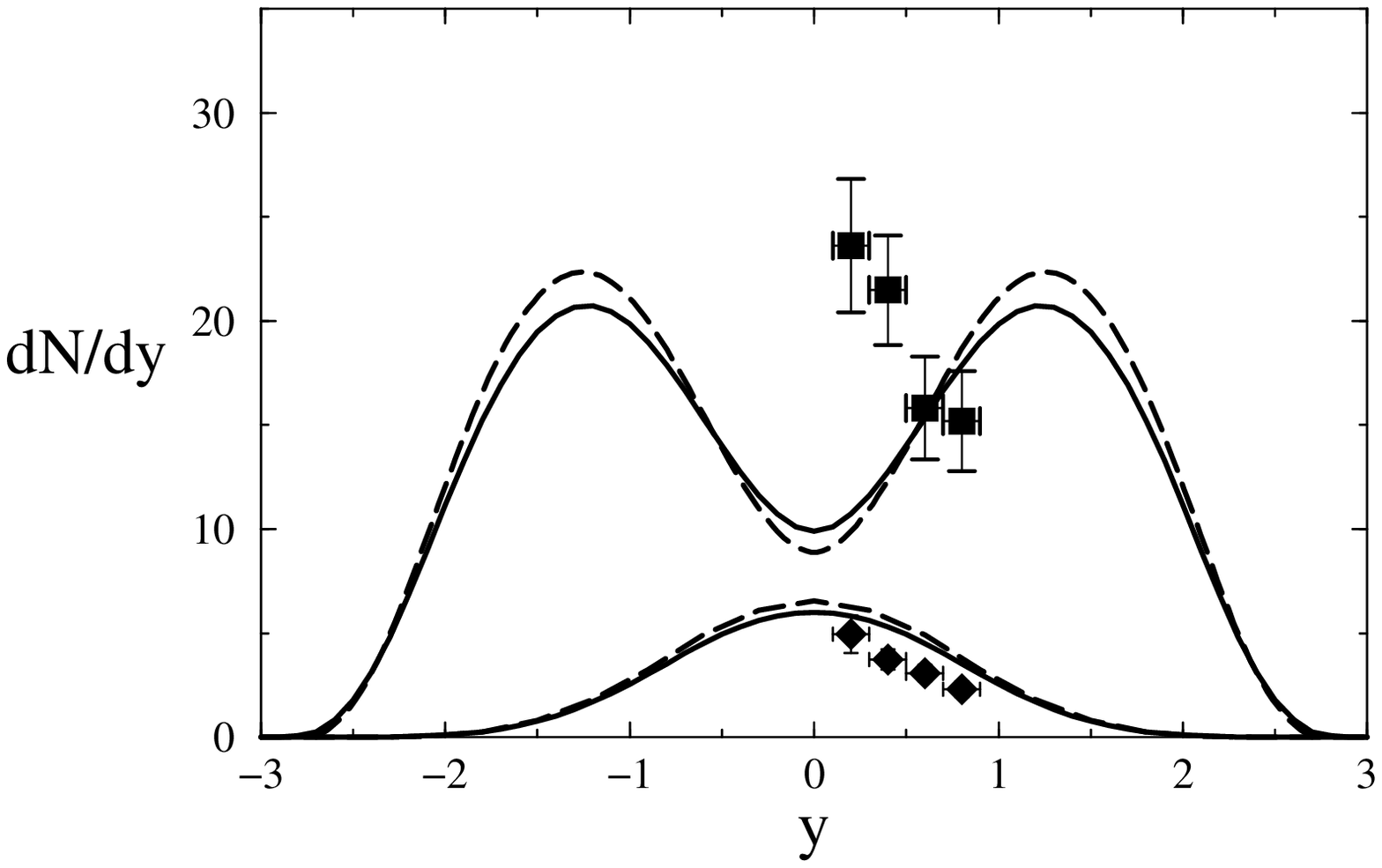}
\caption{Rapidity distribution of $\Lambda$'s and 
$\bar{\Lambda}$'s, solid lines, produced in central 
$Pb\,Pb$ collisions represented in the nucleon-nucleon 
centre-of-mass reference frame. The dashed lines represent
the total hyperons (anti-hyperons) distributions. Data
points are from ref. \cite{NA49L}.} \label{fig:Graph8} 
\end{figure}

\begin{figure}[p]
\epsfbox[0 370 300 720]{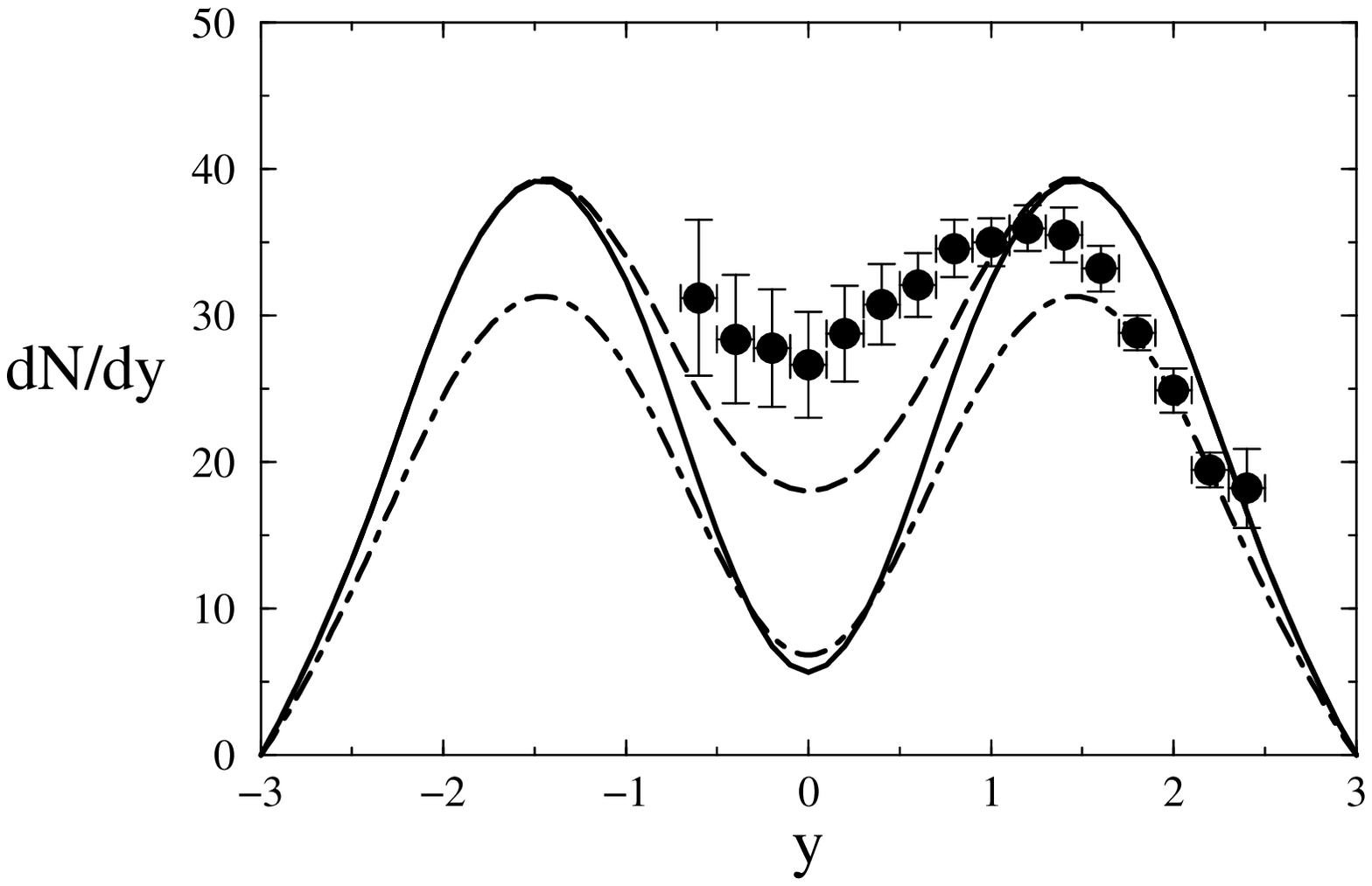}
\caption{Rapidity $p-\bar{p}$ distribution  in central 
$Pb\,Pb$ collisions represented in the nucleon-nucleon 
centre-of-mass reference frame. The solid line corresponds
to the prediction including the correction from hyperon 
electroweak decays into protons applied
to the initial proton distribution, dashed-dotted line. The 
dashed line represents the correction for the excess $K^{+}$ 
over $K^{-}$ (see text for an explanation). Data
points are from ref. \cite{NA49proton}.} \label{fig:Graph9} 
\end{figure}

Figure \ref{fig:Graph8} shows the predictions for the rapidity 
spectra of $\Lambda$ and $\bar{\Lambda}$, solid lines,  produced 
in central Pb-Pb collisions. The experimental points correspond 
to preliminary data from the NA49 Collaboration\citeup{NA49L}. 
The dashed lines correspond to the total hyperon (antihyperon) 
distributions. Figure  \ref{fig:Graph9} compares the result for the
net proton distribution with data from the same collaboration\citeup{NA49proton}. 
These data points were obtained subtracting the background due to 
the decays $\Lambda \rightarrow p \pi^{-}$ using the results of 
a Monte Carlo simulation. Correcting for the difference between the
$\Lambda$ spectrum predicted by this model and the one used in
the experimental analysis, we obtain the solid line out of the
net distribution (dashed-dotted line). I have also corrected 
for the $K^{+}-K^{-}$ asymmetry as explained above, dashed line.
As we can see, the experimental data tend to show a larger 
stopping in Pb-Pb collisions that what the model predicts. 

As mentioned before, I have made use of the Galuber-Gribov 
model to obtain the average numbers of participating 
nucleons and nucleon-nucleon collisions. I have used 
the value $\sigma_{NN}^{\mbox{in}} = 30 \, \mbox{mb}$ 
for the inelastic nucleon-nucleon cross section. The 
profile functions where obtained using the nuclear matter 
distribution given in ref. \cite{NestorPC}. They consist of three parameters, 
individually fitted to nuclear data, 
functions and are more accurate 
than the Saxon-Wood density functions used in the past.
The actual numerical values 
used for central S-S, $\bar{n}_{A} =27.8$ and  $\bar{n} = 62.8$, 
were obtained requiring that $b<1$, where $b$ is the impact 
parameter. For peripheral collisions we have
$\bar{n}_{A} =9.7$ and  $\bar{n} = 14.7$, while for central 
Pb-Pb collisions are $\bar{n}_{A} =204.4$ and  $\bar{n} = 895.9$. 

To perform the computations, I have taken the values of the proton 
and $\Lambda$ transverse mass equal to 1.1 and 1.3 GeV respectively, in 
agreement with the data from \cite{ExpHypHE}.

\section{Conclusions}

In order to explain the data on baryon rapidity spectra presented 
by the NA35 collaboration, I have included into the DPM and the 
QGSM models components that were previously overlooked. They are 
compatible with the $1/N$ expansion of QCD on which these two models are 
based. A discussion over the right structure functions for the string's ends 
leads to the adoption of the ones proposed by the QGSM, 
as they are the only ones consistent with the model's assumptions. 
The fragmentation functions are obtained from the planar 
approximation to Regge diagrams. They have been published 
elsewhere except for the ones corresponding to strange diquark 
that are published here for the first time. All of them are given 
in the appendix. Once taken into account the baryon-number sum-rules, 
the number of free parameters of the model is reduced to 3. The 
numerical values of these parameters have been obtained fitting 
data of inclusive baryon and antibaryon production in proton-proton 
collisions. The sum rules allow us to obtain the 
fragmentation functions that do not contribute 
to proton-proton collisions at ISR energies.

The model has strong theoretical grounds and all the approximations 
done are well justified, being the uncertainties introduced by them 
small and well under control.

At the level of nucleus-nucleus collisions, we have a model which 
contains no free parameters. The $\bar{\Lambda}$ distribution for 
central sulphur-sulphur collisions was computed and a perfect 
agreement with the experimental data was obtained. This result 
does not depend on the new components introduced in this paper; 
all the $\bar{\Lambda}$ particles come from baryon-antibaryon 
formation in the strings, which is essentially independent of 
the parton components at the end of the string. This result is 
very relevant since it clearly shows that the $\bar{s}$ content 
at the central rapidity region is very well understood with no 
need to assume collective effects such as QGP formation.

I have also computed the rapidity distribution for the  $\Lambda$ 
hyperons in central sulphur-sulphur collisions and compared it 
with data from the same collaboration. The results show a remarkable 
increase of the rapidity distribution in the fragmentation region 
well in agreement with the experimental data. This is due to the 
inclusion of chains initiated by diquarks that can contain a quark 
from the nucleon sea. The fact that this sea-quark has a probability 
of being strange, obtained by independent analysis, is 
responsible for this effect. The only disagreement with the data 
is seen at the central rapidity region. 

The net proton distribution for the same collisions has also been 
obtained and compared with data of the same collaboration.
We observe a good agreement between the model results and 
the experimental points in the fragmentation region. Once
the correction of the electric charge asymmetry due to the 
excess of $K^{+}$ over $K^{-}$ is included, we also understand 
the spectrum at mid-rapidities.

The NA35 data on baryon distributions do not support that 
the formation of QGP has taken place or that we need to take
into account collective effects of any sort. I have compare the
model results with the preliminary data on $\Lambda$ and 
$\bar{\Lambda}$ rapidity 
distributions in Pb-Pb collisions form the NA49 Collaboration;  
a partial agreement with the data was found. Unfortunately
there are no points available in the fragmentation region. The 
comparison with the $p-\bar{p}$ spectrum measured by the
same collaboration shows that the model under-predicts the
stopping power in Pb-Pb collisions. This raises the interesting
question about which of the model assumptions fails for the case
of very heavy nuclei. It could be that effects like those of 
references \cite{DB} and \cite{kharzeev} have a significant
contribution to this stopping, falling  within the experimental 
errors in the case of sulphur-sulphur collisions. 
This possibilities should be careful check in hadron-hadron
interactions as they contain the implicit assumption that
the wave function of the colliding hadron couples to the pomeron
over a rapidity region larger than one unit. This may
have an effect on diffractive scattering. The assumption 
that collisions between heavy nucleus can be
decomposed as nearly uncorrelated nucleon-nucleon collisions, 
as implied by the use of the Galuber-Gribov model, may also fail
when the size of the nuclei involved becomes very large. 

Another important conclusion, that has been obtained as 
a byproduct, is that the analysis performed on the 
proton-proton data strongly support a value for the 
intercept of the N trajectory close to $-0.25$. This 
is also a major result of the model that in return 
can help to establish a more accurate value for the 
mass of some resonances.

For future works it is left the detailed study of the 
meson sector. Also the implications of the new structure 
of the pomeron-hadron coupling discussed in the paper 
will be investigated in order to shed light on other 
outstanding problems such as heavy quark formation in 
hadronic and nuclear collisions.

\vspace*{2cm}
{\sl Acknowledgements.}
{\footnotesize 
I want to thank A. Capella
for his continuous support and encouragement, also 
for pointing out a mistake in the first version 
of the manuscript. To him 
and to A. Kaidalov for discussions and many useful 
explanations and comments about their previous works. Also
my gratitude goes to N. Armesto, S. Pepin and M. McDermott for 
comments and discussions and to A. Sabio-Vera for help 
with the technicalities of typesetting the manuscript.
}
\vspace*{0.5cm} 

\section*{Appendix}

I list here the fragmentation functions into baryons and antibaryons. 
The ones not shown can be obtained applying isospin symmetry $p 
\leftrightarrow n$, $u \leftrightarrow d$.

\begin{eqnarray}
D_{1,uu}^{p} &=&  \frac{a_p}{z} z^{2(\alpha_R (0)-\alpha_N (0))} 
(1-z)^{-\alpha_R (0) + \lambda} 
(1+c_{0} z) =  \frac{a_p}{z} z^{1.5} (1+c_{0} z) \nonumber \\ 
D_{1,uu}^{n} &=& \frac{2}{3} \frac{a_p}{z} z^{2(\alpha_R (0)-\alpha_N 
(0))} (1-z)^{-\alpha_R (0) + 
\lambda +2(1-\alpha_R (0))}= \frac{2}{3} \frac{a_p}{z} z^{1.5} (1-z) 
\nonumber \\ 
D_{1,uu}^{\Lambda} &=& \frac{a_{\Lambda}}{z} z^{2(\alpha_R 
(0)-\alpha_N (0))} 
(1-z)^{-\alpha_R (0) + \lambda+ \Delta \alpha +2(1-\alpha_R (0))}  
= \frac{a_{\Lambda}}{z} z^{1.5} (1-z)^{1.5} \nonumber \\ 	
D_{1,ud}^{p} &=& \frac{a_p}{z} z^{2(\alpha_R (0)-\alpha_N (0))} 
(1-z)^{-\alpha_R (0) + \lambda}
= \frac{a_p}{z} z^{1.5} = D_{1,ud}^{n}\nonumber \\ 
D_{1,ud}^{\Lambda} &=& \frac{a_{\Lambda}}{z} z^{2(\alpha_R 
(0)-\alpha_N (0))} 
(1-z)^{-\alpha_R (0)+ \Delta \alpha + \lambda}  
= \frac{a_{\Lambda}}{z} z^{1.5} (1-z)^{0.5} \nonumber \\ 
D_{1,dd}^{p} &=& D_{1,uu}^{n} \quad , \quad 
D_{1,dd}^{n} = D_{1,uu}^{p} \quad , \quad 
D_{1,dd}^{\Lambda} = D_{1,uu}^{\Lambda}\nonumber \\ 
D_{1,us}^{p} &=& \frac{1}{2} \frac{a_p}{z} z^{2(\alpha_R (0)-\alpha_N 
(0))} 
(1-z)^{-\alpha_R (0) + \lambda+ \Delta \alpha +2(1-\alpha_R (0))} 
(1+z^{\Delta \alpha }) = D_{1,us}^{n} \nonumber \\
&=& \frac{1}{2} \frac{a_p}{z} z^{1.5} (1-z)^{1.5}  
(1+z^{\frac{1}{2}}) \nonumber \\
D_{1,us}^{\Lambda} &=& \frac{1}{2} \frac{a_{\Lambda}}{z} 
z^{2(\alpha_R (0)-\alpha_N (0))} 
(1-z)^{-\alpha_R (0) + \lambda}  (1+z^{\Delta \alpha}) (1+c_1 z) 
\nonumber \\
&=& \frac{1}{2} \frac{a_{\Lambda}}{z} z^{1.5} (1+z^{ \frac{1}{2}}) 
(1+c_1 z) \nonumber \\
D_{1,ds}^{p} &=& D_{1,us}^{p} \quad , \quad 
D_{1,ds}^{n} = D_{1,us}^{n}  \quad , \quad 
D_{1,ds}^{\Lambda} = D_{1,us}^{\Lambda} \quad , \quad 
D_{1,ss}^{p} = 0 \quad , \quad 
D_{1,ss}^{n} = 0 \nonumber \\
D_{1,ss}^{\Lambda} &=& \frac{a_{\Lambda}}{z} z^{2(\alpha_R 
(0)-\alpha_N (0))+\Delta \alpha} 
(1-z)^{-\alpha_R (0) + \lambda + \Delta \alpha +2(1-\alpha_R (0))} 
(1+c_2 z)  \nonumber \\
&=& \frac{a_{\Lambda}}{z} z^{2}(1-z)^{2} (1+c_2 z)  \nonumber \\
%
%
D_{2,uu}^{p,n} &=&  D_{2,ud}^{p,n} = D_{2,dd}^{p,n} = 
\frac{a_{\bar{p}}}{z} (1-z)^{-\alpha_R (0) + \lambda + 4(1-\alpha_N 
(0))} =  \frac{a_{\bar{p}}}{z} (1-z)^5 \nonumber \\
D_{2,uu}^{\Lambda} &=& D_{2,ud}^{\Lambda} = D_{2,dd}^{\Lambda} 
=\frac{a_{\bar{\Lambda}}}{z}(1-z)^{-\alpha_R (0) + \lambda +\Delta 
\alpha +  4(1-\alpha_N (0))}= \frac{a_{\bar{\Lambda}}}{z}(1-z)^{5.5}  
\nonumber \\
D_{2,us}^{p,n}  &=& D_{2,ds}^{p,n} = \frac{a_{\bar{p}}}{z} 
(1-z)^{-\alpha_R (0) + \lambda +\Delta \alpha +  4(1-\alpha_N (0))} 
=  \frac{a_{\bar{p}}}{z} (1-z)^{5.5} \nonumber \\
D_{2,us}^{\Lambda}  &=& D_{2,ds}^{\Lambda} =  
\frac{a_{\bar{\Lambda}}}{z} (1-z)^{-\alpha_R (0) + \lambda + 2 \Delta 
\alpha +  4(1-\alpha_N (0))} 
= \frac{a_{\bar{\Lambda}}}{z}(1-z)^{6}  \nonumber \\
D_{2,ss}^{p,n}  &=&  \frac{a_{\bar{p}}}{z} (1-z)^{-\alpha_R (0) + 
\lambda + 2 \Delta \alpha +  4(1-\alpha_N (0))}
=  \frac{a_{\bar{p}}}{z} (1-z)^{6} \nonumber \\
D_{2,ss}^{\Lambda}  &=&  \frac{a_{\bar{\Lambda}}}{z} (1-z)^{-\alpha_R 
(0) + \lambda + 3 \Delta \alpha +  4(1-\alpha_N (0))}
= \frac{a_{\bar{\Lambda}}}{z}(1-z)^{6.5}  \nonumber  \\
%
%
D_{u}^{p}  &=& D_{d}^{n} =  \frac{a_{\bar{p}}}{z} (1-z)^{\alpha_R 
(0)-2 \alpha_N (0) + \lambda }  =  \frac{a_{\bar{p}}}{z} 
(1-z)^{1.5}   \nonumber  \\
D_{u}^{\Lambda}  &=& D_{d}^{\Lambda} =  \frac{a_{\bar{\Lambda}}}{z} 
(1-z)^{\alpha_R (0)-2 \alpha_N (0) + \lambda + \Delta \alpha } =  
\frac{a_{\bar{\Lambda}}}{z} (1-z)^{2}   \nonumber  \\
D_{d}^{p}  &=& D_{u}^{n} =  \frac{a_{\bar{p}}}{z} (1-z)^{\alpha_R 
(0)-2 \alpha_N (0) + \lambda }  ( \frac{1}{3} +  \frac{2}{3}(1-z))
=  \frac{a_{\bar{p}}}{z} (1-z)^{1.5}   ( \frac{1}{3} +  
\frac{2}{3}(1-z))  \nonumber  \\
D_{s}^{p,n}  &=&  \frac{a_{\bar{p}}}{z} (1-z)^{\alpha_R (0)-2 
\alpha_N (0) + \lambda + 2 (1-\alpha_R (0)) + \Delta \alpha } 
=  \frac{a_{\bar{p}}}{z} (1-z)^{3}   \nonumber  \\
D_{s}^{\Lambda}  &=&  \frac{a_{\bar{\Lambda}}}{z} (1-z)^{\alpha_R 
(0)-2 \alpha_N (0) + \lambda }  
=  \frac{a_{\bar{\Lambda}}}{z} (1-z)^{1.5}   \nonumber  \\
%
%
D_{uu}^{\bar{p},\bar{n}} &=&  D_{ud}^{\bar{p} , \bar{n}} = 
D_{dd}^{\bar{p},\bar{n}}  
=  \frac{a_{\bar{p}}}{z} (1-z)^{\alpha_R (0) - 2 \alpha_N (0) + 
\lambda +  2(1-\alpha_N (0))} 
=  \frac{a_{\bar{p}}}{z} (1-z)^{4}  \nonumber \\
D_{uu}^{\bar{\Lambda}} &=& D_{ud}^{\bar{\Lambda}}= 
D_{dd}^{\bar{\Lambda}}   
= \frac{a_{\bar{\Lambda}}}{z}(1-z)^{\alpha_R (0) - 2 \alpha_N (0) + 
\lambda + \Delta \alpha +  2(1-\alpha_N (0))} 
=  \frac{a_{\bar{\Lambda}}}{z}(1-z)^{4.5}  \nonumber \\
D_{us}^{\bar{p},\bar{n}} &=&  D_{ds}^{\bar{p} , \bar{n}} =  
=  \frac{a_{\bar{p}}}{z} (1-z)^{\alpha_R (0) - 2 \alpha_N (0) + 
\lambda + \Delta \alpha  +  2(1-\alpha_N (0))}
=  \frac{a_{\bar{p}}}{z} (1-z)^{4.5}  \nonumber \\
D_{us}^{\bar{\Lambda}} &=&D_{ds}^{\bar{\Lambda}}  
= \frac{a_{\bar{\Lambda}}}{z}(1-z)^{\alpha_R (0) - 2 \alpha_N (0) + 
\lambda + 2 \Delta \alpha +  2(1-\alpha_N (0))} 
=  \frac{a_{\bar{\Lambda}}}{z}(1-z)^{5}  \nonumber \\
D_{ss}^{\bar{p},\bar{n}} &=& \frac{a_{\bar{p}}}{z} (1-z)^{\alpha_R 
(0) - 2 \alpha_N (0) + \lambda + 2 \Delta \alpha  +  2(1-\alpha_N 
(0))} 
=  \frac{a_{\bar{p}}}{z} (1-z)^{5}  \nonumber \\
D_{ss}^{\bar{\Lambda}} &=&  
\frac{a_{\bar{\Lambda}}}{z}(1-z)^{\alpha_R (0) - 2 \alpha_N (0) + 
\lambda + 3 \Delta \alpha +  2(1-\alpha_N (0)) } 
=  \frac{a_{\bar{\Lambda}}}{z}(1-z)^{5.5}  \nonumber \\
D_{u}^{\bar{p},\bar{n}} &=&  D_{d}^{\bar{p} , \bar{n}} =  
\frac{a_{\bar{p}}}{z} (1-z)^{- \alpha_R (0) + \lambda +  2(1-\alpha_N 
(0))}  
=  \frac{a_{\bar{p}}}{z}(1-z)^{2.5}  \nonumber \\
D_{u}^{\bar{\Lambda}} &=&   \frac{a_{\bar{\Lambda}}}{z}(1-z)^{- 
\alpha_R (0) + \lambda + \Delta \alpha + 2(1-\alpha_N (0))}  
=  \frac{a_{\bar{\Lambda}}}{z}(1-z)^{3}  \nonumber \\
D_{s}^{\bar{p},\bar{n}} &=&  \frac{a_{\bar{p}}}{z} (1-z)^{- \alpha_R 
(0) + \lambda + \Delta \alpha +  2(1-\alpha_N (0))}  
=  \frac{a_{\bar{p}}}{z}(1-z)^{3}  \nonumber \\
D_{s}^{\bar{\Lambda}} &=&   \frac{a_{\bar{\Lambda}}}{z}(1-z)^{- 
\alpha_R (0) + \lambda + 2 \Delta \alpha + 2(1-\alpha_N (0))}  
=  \frac{a_{\bar{\Lambda}}}{z}(1-z)^{3.5}  \nonumber 
\end{eqnarray}

\end{document}